\documentclass{aa}  

\usepackage{graphicx}
\usepackage{pifont}
\usepackage[colorlinks=true,citecolor=blue]{hyperref}
\usepackage[export]{adjustbox}
\usepackage{txfonts}

\begin{document} 

   \title{Introducing the AIDA-TNG project:\\Galaxy formation in alternative dark matter models}
    \titlerunning{Introducing the AIDA-TNG simulations}
    \authorrunning{Despali et al.}

   \author{Giulia Despali
          \inst{1,2,3}\fnmsep\thanks{giulia.despali@unibo.it},
          Lauro Moscardini\inst{1,2,3},
          Dylan Nelson\inst{4},
          Annalisa Pillepich\inst{5}, \\
          Volker Springel\inst{6},
          Mark Vogelsberger\inst{7}
          }

   \institute{Dipartimento di Fisica e Astronomia "Augusto Righi", Alma Mater Studiorum Università di Bologna, via Gobetti 93/2, I-40129 Bologna, Italy \label{1}
   \and INAF-Osservatorio di Astrofisica e Scienza dello Spazio di Bologna, Via Piero Gobetti 93/3, I-40129 Bologna, Italy \label{2}
   \and INFN-Sezione di Bologna, Viale Berti Pichat 6/2, I-40127 Bologna, Italy \label{3}
   \and Universität Heidelberg, Zentrum für Astronomie, ITA, Albert-Ueberle-Str. 2, 69120 Heidelberg, Germany
   \label{4}
   \and Max-Planck-Institut für Astronomie, Königstuhl 17, 69117 Heidelberg, Germany \label{5}
      \and Max-Planck-Institut für Astrophysik, Karl-Schwarzschild-Straße 1, D-85740, Garching bei München, Germany \label{6}
      \and Department of Physics, Kavli Institute for Astrophysics and Space Research, Massachusetts Institute of Technology, Cambridge, MA 02139, USA \label{7}
        }

   \date{}

  \abstract{
  We introduce the AIDA-TNG project, a suite of cosmological magnetohydrodynamic simulations that simultaneously model galaxy formation and different variations in the underlying dark matter model. We consider the standard cold dark matter model and five variations, including three warm dark matter scenarios and two self-interacting models with a constant or velocity-dependent cross-section. In each model, we simulated two cosmological boxes of 51.7 and 110.7 Mpc on a side with the same initial conditions as TNG50 and TNG100, and we combined the variations in the physics of dark matter with the fiducial IllustrisTNG galaxy formation model. The AIDA-TNG runs are thus ideal for studying the simultaneous effects of baryons and alternative dark matter models on observable properties of galaxies and large-scale structures. We resolved haloes in the range between $10^{8}$ and $4\times10^{14}\,$M$_{\odot}$ and scales down to the nominal resolution of 570 pc in the highest-resolution runs. This work presents the first results on statistical quantities such as the halo mass function and the matter power spectrum. We quantified the modification in the number of haloes and the power on scales smaller than 1 Mpc due to the combination of baryonic and dark matter physics. Despite being calibrated on cold dark matter, we find that the TNG galaxy formation model can produce a realistic galaxy population in all scenarios. The stellar and gas mass fraction, stellar mass function, black hole mass as a function of stellar mass, and star formation rate density are very similar in all dark matter models, with some deviations only in the most extreme warm dark matter model. Finally, we also quantify changes in halo structure due to warm and self-interacting dark matter, which appear in the density profiles, concentration-mass relation, and galaxy sizes.  
  }

   \keywords{Cosmology: dark matter -- Cosmology: large-scale structure of Universe -- Galaxies -- Galaxies: evolution}

   \maketitle


\section{Introduction}

One of the foundations of the concordance cosmological model is that approximately 85\% of the matter content of the Universe is in the form of some yet unknown component that can only be detected through its gravitational effect. Namely, this component is dark matter. Starting from small density perturbations in the early Universe, dark matter leads the gravitational collapse on large scales, forming the cosmic-web structure of filaments. Along these filaments, gas flows to the potential wells of dark matter haloes, where it cools and creates stars -- eventually leading to the formation of galaxies. Despite decades of experimental searches, the favoured particle candidates for cold dark matter (CDM) -- axions and weakly interacting massive particles (WIMPS) -- have so far not been detected. This has motivated the study of alternative models that are allowed by particle physics assumptions and span a wide range of masses -- from the electronvolt to the gigaelectron volt scale.

Moreover, while the CDM model is very successful at explaining the large-scale structure distribution of the Universe \citep{planck1_14}, it has been challenged by observations at the scale of galaxies and below \citep{bullock17}. In the past few years, inconsistencies between CDM numerical simulations and observations that led to the famous `missing satellites problem' \citep{klypin99} have been mitigated with the inclusion of baryonic effects in simulations \citep{brooks13} and with the discovery of new faint satellites in the Milky Way \citep{bechtol2015,drlica2015,kim18}. However, other possible tensions have arisen, such as the `core-cusp' and `too-big-to-fail' problems \citep{boylan-kolchin11} or the diversity of rotation curves \citep{kaplinghat20}. Recent observational analyses have also identified new challenges for CDM. For example, the lensing properties of subhaloes appear inconsistent with CDM both at galaxy \citep{minor17,ballard24,despali24,enzi24} and cluster scales \citep{meneghetti20,ragagnin22,meneghetti23}. In both cases, the observed lensing signal is inferred to be stronger than predicted, and the subhalo concentrations required to explain it are outliers within the CDM model. It is thus still uncertain if simulated CDM predictions can be brought completely into agreement with observations by modifications of baryonic feedback and other aspects of the subgrid physics in simulations or if these discrepancies point towards new physics in the dark sector. 

Many previous works have addressed the aforementioned tensions, with some developing a theoretical and computational framework especially suited for warm dark matter (WDM) and self-interacting dark matter (SIDM). These models were originally motivated by models of the satellite population of galaxies (for WDM), the profiles of dwarf galaxies, and merging clusters (for SIDM). While these effects are certainly among the most relevant, the most stringent constraints on the WDM particle mass are derived from observations of the Lyman-$\alpha$ forest \citep{irsic24} and flux-ratio anomalies of lensed quasars \citep{gilman19}, while the self-interaction cross-section is strongly constrained at the scale of massive elliptical galaxies and galaxy clusters \citep{meneghetti01,peter13,despali19,despali22,eckert22,shen22,mastromarino23}. In general, any dark matter model that aspires to explain observations should be challenged with reproducing the halo and galaxy properties spanning many decades in mass and not only in a narrow range of scales. This is the fundamental aim of the AIDA-TNG project.

\begin{figure}
\centering
   \includegraphics[width=\columnwidth]{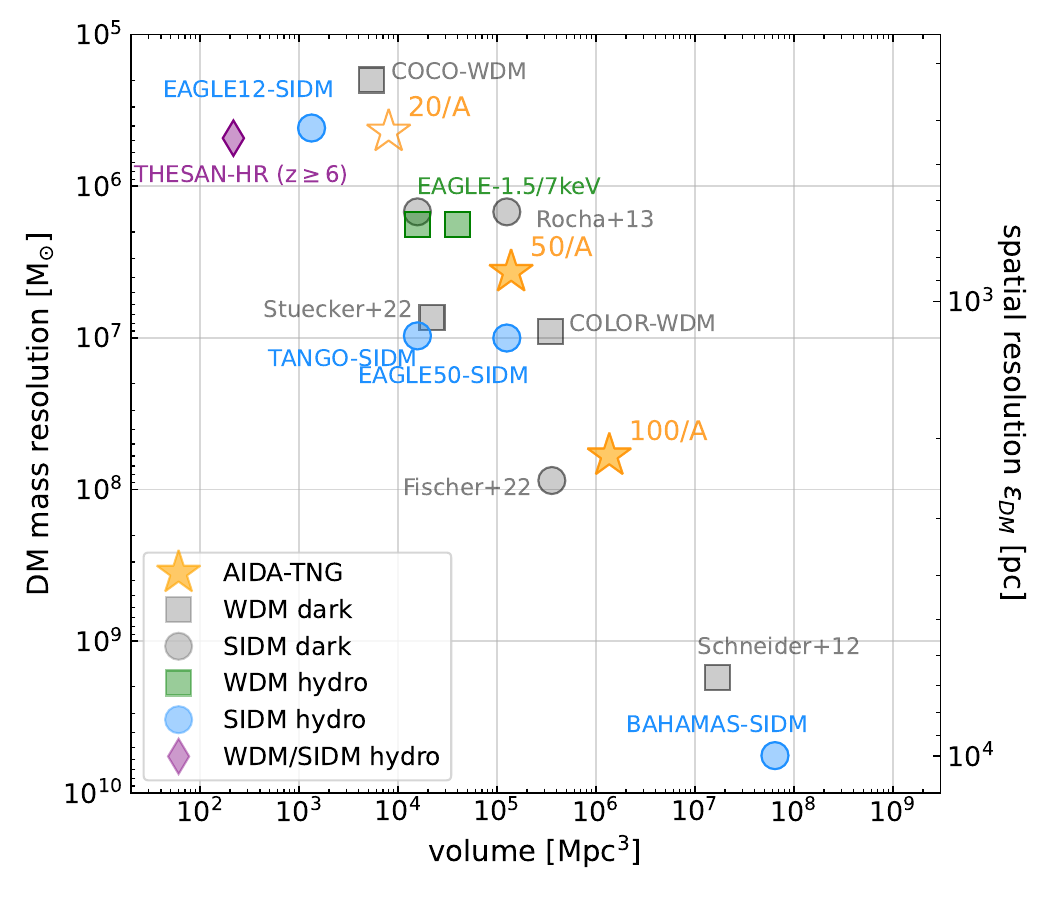}
   \caption{Available cosmological simulations for ADM models, compared in volume ($x$-axis) and resolution ($y$-axis), to the AIDA-TNG flagship runs (yellow stars) introduced in this paper. Grey symbols show DMO runs \citep{schneider12,rocha13,bose16,stuecker22,fischer22,fischer24}, while the simulations including baryonic physics are shown in colour \citep{robertson18,robertson19,forouhar22,shen24,correa24,oman24}. The present work introduces the 50 and 100 Mpc AIDA runs, while the highest-resolution 20 Mpc box will be presented in an upcoming paper.}
    \label{fig:sims}
\end{figure}

The majority of the work in the field of alternative dark matter is based on zoom-in simulations of limited samples of objects \citep{zavala13,rocha13,fry15,robles17,creasey17,despali19,despali22,bhattacharyya22,nadler23,oneil23,ragagnin24}. These simulations can have the advantage of high spatial and mass resolution, but they lack the required statistics to test models on multiple scales and fully understand potential biases in the sample selection. At the same time, the inclusion of baryonic physics is fundamental for studying small scales, but only a limited number of available alternative dark matter (ADM) hydrodynamical simulations are of cosmological scale (see Fig.~\ref{fig:sims}). For WDM, \citet{oman24} recently presented two cosmological runs that use the Eagle galaxy formation model: one 25 Mpc box where the WDM model considers a 1.5 keV thermal relic candidate and one 34 Mpc box reproducing the effect of 7.1 keV sterile neutrinos. In SIDM, the EAGLE-SIDM \citep{robertson18,robertson21,forouhar22} and BAHAMAS-SIDM \citep{robertson19} simulations have been the first to include both self-interactions and baryonic physics on cosmological scales. This set includes elastic self-interactions with constant cross-sections $\sigma/m_{\chi}=[0.1,1]\,{\rm cm^{2}g^{-1}}$ and a model with a mild velocity dependence peaking at $\sigma/m_{\chi}=3\,{\rm  cm^{2}g^{-1}}$. More recently, the TANGO-SIDM simulations \citep{correa22,correa24} explored velocity-dependent SIDM models in 25 Mpc volumes, including the SWIFT-EAGLE physics model \citep{schaller24,bahe22,borrow23}. Finally, the THESAN-HR \citep{shen24} simulations include different dark matter variations; however, given their focus on the epoch of reionisation, they only follow a $\sim$6 Mpc box until $z=6$. 

The AIDA-TNG project advances the field of ADM simulations by providing a consistent set of high-resolution cosmological boxes simulated with the same galaxy formation model and multiple WDM and SIDM models, in addition to CDM. We provide a large simulated sample of galaxies and break the degeneracy between baryonic and dark matter effects at small scales. In this work, we present simulations evolving two cosmological volumes of 51.7 and 110.7 Mpc on a side evolved with the \textsc{AREPO} code \citep{springel10} in different dark matter scenarios and in the TNG galaxy formation model \citep{weinberger17,pillepich18}.

We present the first results from the AIDA-TNG suite, providing an overview of the multiple effects of ADM models. In Sect.~\ref{sec:sims}, we describe the set-up of the AIDA simulations and discuss the considered dark matter models and their properties. We then present our main results in Sects.~\ref{sec:massf} to \ref{sec:power}. We start by discussing the halo and stellar mass function in Sect. \ref{sec:massf} and their evolution with redshift. Section~\ref{sec:prof} focuses on the effects of ADM on the distribution of dark matter and baryons within haloes, exploring the density profiles and concentration-mass relation. We then look at the properties of galaxies in Sect.~\ref{sec:galaxies}. Finally, Sect.~\ref{sec:power} presents measurements of the matter power spectra of the different mass components and their evolution in time. In Sect.~\ref{sec:conc}, we summarise our results and present our conclusions.


\section{The AIDA simulations} \label{sec:sims}

{\renewcommand{\arraystretch}{1.3}
\begin{table*}
\caption{AIDA-TNG runs presented in this work.}
\label{table:1}
\centering
\begin{tabular}{c c c c c c c| c c c | c c }
\hline\hline
Name & Box & Physics & $m_{\rm DM}$ & $m_{\rm bar}$ & $\epsilon_{\rm DM,*}^{z=0}$ & CDM & WDM  & WDM  & WDM& SIDM  & vSIDM \\
& $\rm{[Mpc]}$&  & $ [\rm{M}_{\odot}$] & $ [\rm{M}_{\odot}$] & [kpc] & & 1\,keV & 3\,keV & 5\,keV & 1$\,\rm{cm^{2}g^{-1}}$ & Correa+21 \\
\hline\hline
        
   100/A& 110.7 & DMO & 7.1$\times 10^{7}$ & - & 1.48 &\ding{51} & - &\ding{51} & - &\ding{51}  &\ding{51} \\
   & & FP & 6.0$\times 10^{7}$ & 1.1$\times 10^{7}$ & 1.48 & \ding{51} & - & \ding{51} & -& \ding{51}  & \ding{51} \\
  100/B & 110.7 & DMO & 5.7$\times 10^{8}$ & -  & 2.95 &\ding{51} & \ding{51} & \ding{51} & - & \ding{51}  & \ding{51} \\
    & & FP & 4.8$\times 10^{8}$ & 8.9$\times 10^{7}$ & 2.95& \ding{51} & \ding{51} & - & -& \ding{51}  & \ding{51} \\
    
    \hline
   50/A &51.7 & DMO & 4.3$\times 10^{6}$ & - & 0.57 & \ding{51} & - & \ding{51} & \ding{51} & \ding{51}  & \ding{51} \\
   & & FP & 3.6$\times 10^{6}$ & 6.8$\times 10^{5}$ & 0.57 & \ding{51} & - &  \ding{51} & -&\ding{51}  &\ding{51} \\
   50/B & 51.7 & DMO & 3.4$\times 10^{7}$ & - & 1.15 & \ding{51} & \ding{51} & \ding{51} & -& \ding{51}  & \ding{51} \\
   & & FP & 2.9$\times 10^{7}$ & 5.4$\times 10^{6}$ & 1.15 & \ding{51} & \ding{51} & \ding{51}& - & \ding{51}  & \ding{51} \\
\hline\hline
\end{tabular}
\tablefoot{ We run each box at two resolution levels, where `A' is the highest. The 100/A(B) and 50/A(B) boxes start from the same initial conditions of the TNG100-2(3) and TNG50-2(3) boxes of the original IllustrisTNG project \citep{nelson19}. We list the resolution of each run and mark the dark matter models that are included in this work.}
\end{table*}

}

The AIDA-TNG simulations, \emph{Alternative Dark Matter in the TNG universe} (AIDA for short, hereafter), consist of three cosmological volumes of increasing resolution and decreasing size simulated in cold and ADM models, with and without baryonic physics. In this work we present the two largest boxes of 110.7 and 51.7 Mpc on a side and the first results obtained from their analysis, while a smaller (and higher resolution) 20 Mpc box will be introduced in an upcoming work. The uniqueness of the AIDA set lies in the fact that we run the same cosmological volumes in multiple dark matter scenarios, spanning cold, warm and self-interacting dark matter. Moreover, we create both a dark matter-only (DMO) and a full-physics (FP) version of each run so that we can disentangle the effects of ADM and baryonic physics as described by the IllustrisTNG galaxy formation model \citep[TNG hereafter][]{weinberger17,pillepich18}. We discuss the considered dark matter scenarios and the integration with the TNG model in more detail in Sects.~\ref{subsec:wdm}, \ref{subsec:sidm} and \ref{subsec:tng}.

All simulations start at redshift $z=127$ and adopt the cosmological model from \citet{planckxxiv}: $\Omega_{\rm m}=0.3089$, $\Omega_{\rm \Lambda}=0.6911$, $\Omega_{\rm b}=0.0486$, $H_{0}=0.6774$ and $\sigma_{8}=0.8159$. The initial conditions (ICs) have been generated with \textsc{N-GenIC} code \citep{springel05a} by applying the Zel’dovich approximation on a glass distribution of particles with a linear matter transfer function computed using the \textsc{CAMB} code \citep{camb}. The 110.7 and 51.7 Mpc boxes (corresponding to 75 and 35 $h^{-1}{\rm Mpc}$, respectively) are run from the same ICs of the corresponding runs of the original IllustrisTNG simulations \citep{pillepich18b}. For each case, we simulate two resolution levels: `A' corresponds to the TNG100-2 and TNG50-2 runs, while `B' corresponds to the TNG100-3 and TNG50-3 runs. Therefore, we can compare to many previous results on galaxy formation in CDM and better understand systematics. Haloes and subhaloes are identified using the \textsc{SUBFIND} algorithm, producing catalogues of the halo and galaxy properties with the same information available for the TNG runs \citep{nelson19}. In the FP 50/A runs, we reach a best spatial resolution of $\epsilon_{\rm DM,*}=570$ parsecs (gravitational softening length), as well as a dark matter and target baryon mass resolution of m$_{\rm DM}=3.6\times 10^{6}\,{\rm M}_{\odot}$ and m$_{\rm bar}=6.8\times 10^{5}\,{\rm M}_{\odot}$, respectively. 

Table~\ref{table:1} lists the properties of all other runs, and which combinations of box size, resolution and dark matter model have been computed. The choice was made with the aim of a economic use of computational resources and is based on which models produce significant differences compared to CDM at a given resolution -- an issue particularly relevant for the WDM models, as discussed in the next section.

\begin{figure*}
\centering
   \includegraphics[width=0.9\columnwidth]{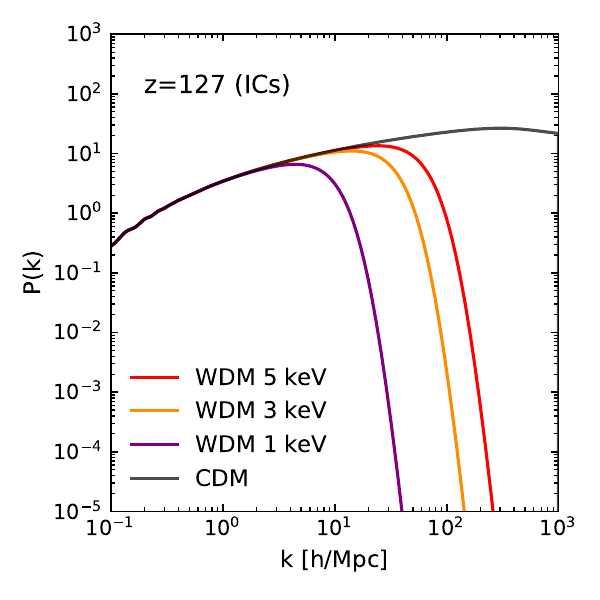}
   \includegraphics[width=0.9\columnwidth]{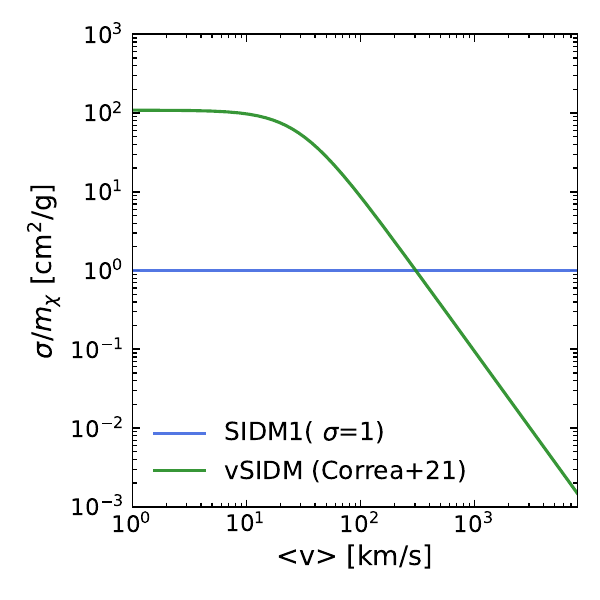}
   \caption{Properties that distinguish alternative models from CDM. {\it Left:} Input matter power spectrum $P(k)$ in CDM and WDM models at the initial time of the simulations ($z=127$). Warmer models show a cut-off at increasingly larger scales, corresponding to smaller $k$ values. {\it Right:} Self-interaction cross-section $\sigma/m_{\chi}$ as a function of velocity. We consider a model with a constant cross-section (blue line) and one with a steep velocity dependence (green line) from \citet{correa21}.
   }
              \label{fig:models}%
\end{figure*}

\subsection{Warm dark matter models} \label{subsec:wdm}

In WDM models, dark matter particles are lighter (keV) than CDM candidates (GeV) and thus become non-relativistic at sufficiently late times so as to suppress, by free streaming, the formation of low-mass galactic-scale haloes with respect to CDM. The number of satellites of the Milky Way and the properties of its stellar streams, strong gravitational lensing and the Lyman-$\alpha$ forest are all well-known probes of the WDM particle mass $m_{\rm WDM}$. Each of these is sensitive to the number of low-mass haloes and subhaloes in certain redshift and halo mass ranges: our Galaxy, gravitational lenses at distances $0.1\leq z_{\rm lens}\leq3$, and the distribution of neutral gas on large scales. Joint constraints that combine the results of such methods are thus more robust than individual measurements. Very warm models are currently already excluded, with upper limits at m$_{\rm WDM}\geq 5\,{\rm  keV}$ \citep{gilman20}, m$_{\rm WDM}\geq 6.046\,{\rm keV}$ \citep{enzi21}, m$_{\rm WDM}\geq 7.4\, {\rm keV}$ \citep{nadler21}, m$_{\rm WDM}\geq 4.1\,{\rm keV}$ \citep{irsic24}: If dark matter is warm, it still needs to be relatively cold. 

In practice, a suppressed power spectrum is used to re-create the initial conditions with the \textsc{N-GenIC} code \citep{springel05b,angulo12}. The difference between WDM and CDM, and thus the amount of suppression, can be fully modelled by a transfer function \citep{bode01}:
\begin{eqnarray}
    \frac{P(k)_{\rm WDM}}{P(k)_{\rm CDM}} & = & |T_{k}|^{2},\\
    \mathrm{where} \qquad T_{k} & = & \left[1+(\alpha k)^{2\nu}\right]^{-5/\nu}, \nonumber\\
    \mathrm{and} \qquad  \alpha & = & 0.048\left(\frac{\Omega_{X}}{0.4}\right)^{0.15}\left(\frac{h}{0.65}\right)^{1.3}\left(\frac{\mathrm{keV}}{m_{X}}\right)^{1.15}\left(\frac{1.5}{g_{X}}\right)^{0.29}. \nonumber
\end{eqnarray}
The parameter $\alpha$ characterises the scale of damping and $\nu=1.2$, $g_{X}=1.5$ \citep{viel05}. In this way, small-scale fluctuations (high $k$) are erased while the power spectrum is identical to the CDM one on large scales. From the ratio of the WDM and CDM $P(k)$, one can define the half-mode mass $M_{\rm hm}$ that corresponds to the scale at which the power spectrum is suppressed by one half and is thus the halo mass below which we expect to erase almost all structures. In the AIDA sample, we consider three different WDM models where the mass of the dark matter particle candidates is m$_{\rm WDM}=(1,3,5)\,{\rm keV}$, corresponding to the half-mode masses M$_{\rm hm}=(1.26\times10^{10}, 3.26\times10^{8},5.95\times10^{7})\, h^{-1}{\rm M}_{\odot}$. The left panel of  Fig. \ref{fig:models} shows the initial power spectra of the considered models. In comparison to the existing observational limits:
\begin{itemize}
    \item The warmest model (WDM1 -- m$_{\rm WDM}=1\,{\rm keV}$) is already ruled out since it is incompatible with structure formation constraints. However, we use it as an extreme case to emphasise any differences with respect to the CDM simulations and to test the effects of baryonic physics in such a different scenario. We do not run this model at the highest resolution given that it erases most structures with masses M$_{\rm vir}\sim 10^{10}\,{\rm M}_{\odot}$ (see Sect. \ref{sec:massf}). These are already well sampled by the mass resolution of the 100/B and 50/B boxes (see Table~\ref{table:1}) which are sufficient to study it properly. 
    \item The intermediate scenario (WDM3 -- m$_{\rm WDM}=3\,{\rm keV}$), is somewhat at the edge of observational constraints: it is already ruled out by combined analysis, while some individual probes (for example gravitational lensing) still allow it. Given the lack of cosmological hydrodynamical simulations with this model, most constraints are based on predictions from DMO runs. We may not yet fully understand possible systematics coming from zoom-in runs of small samples of galaxies and the extent of possible deviations from CDM at all scales, making it an interesting scenario. Moreover, this model (together with its 7 keV sterile-neutrino counterpart) has been used in the majority of recent numerical explorations of WDM effects \citep{lovell12,lovell14,oman24}, and thus it is useful for comparisons to previous work. 
    \item Finally, the coldest model (WDM5 -- m$_{\rm WDM}=5\,{\rm keV}$) is still considered a viable alternative to CDM by observational constraints. It starts to suppress structures significantly only in the regime of M$\leq5\times10^{8}\,{\rm M}_{\odot}$, and thus we only consider it in the highest-resolution runs, which start to probe this mass range. The results presented in Sect.~\ref{sec:massf} demonstrate that this model is very close to CDM, given the maximum mass and spatial resolution of the two large boxes. Therefore, we choose not to create a FP version of the 50/A WDM5 box, which would be a waste of computational resources, and postpone such a combination to the upcoming higher-resolution 20 Mpc simulations.
\end{itemize}

The analysis of WDM simulations requires a careful treatment of some numerical effects. It has been demonstrated that any model with a steep suppression in the power spectrum (such as WDM) can be affected by the formation of spurious structures at the low-mass end caused by artificial fragmentation. This happens at a limiting mass M$_{\rm lim}=10.1\,\bar{\rho}_{m}\left(L_{\rm box}/N_{\rm dm}^{1/3}\right)k_{p}^{-2}$, where $k_{p}$ corresponds to the maximum in the dimensionless power-spectrum $\Delta^{2}(k)$ \citep{wang07c}. Thus, M$_{\rm lim}$ depends on the simulation size $L_{\rm box}$ and resolution of the dark matter component N$_{\rm dm}$, and on the WDM particle mass. We calculated M$_{\rm lim}$ for each run and found that its values are either below the smallest haloes that we considered or nicely correspond to the upturn in the halo mass function (see the discussion in Sect. \ref{sec:massf}). We thus used M$_{\rm lim}$ as a lower limit for the mass function calculation and do not attempt, at this stage, to identify the spurious haloes and separate them from the real ones. \citet{lovell14} proposed an empirical way of eliminating spurious subhaloes by tracking their particles to the initial conditions and measuring the shapes of the Lagrangian regions occupied by the protohalo. They identified a typical sphericity of CDM protohaloes and used it to select only WDM systems of comparable shape, disregarding those with very elongated protohaloes. We will apply this method in future works focused on the halo and subhalo properties at the low-mass end.

\begin{figure*}
    \centering
   \includegraphics[width=\textwidth]{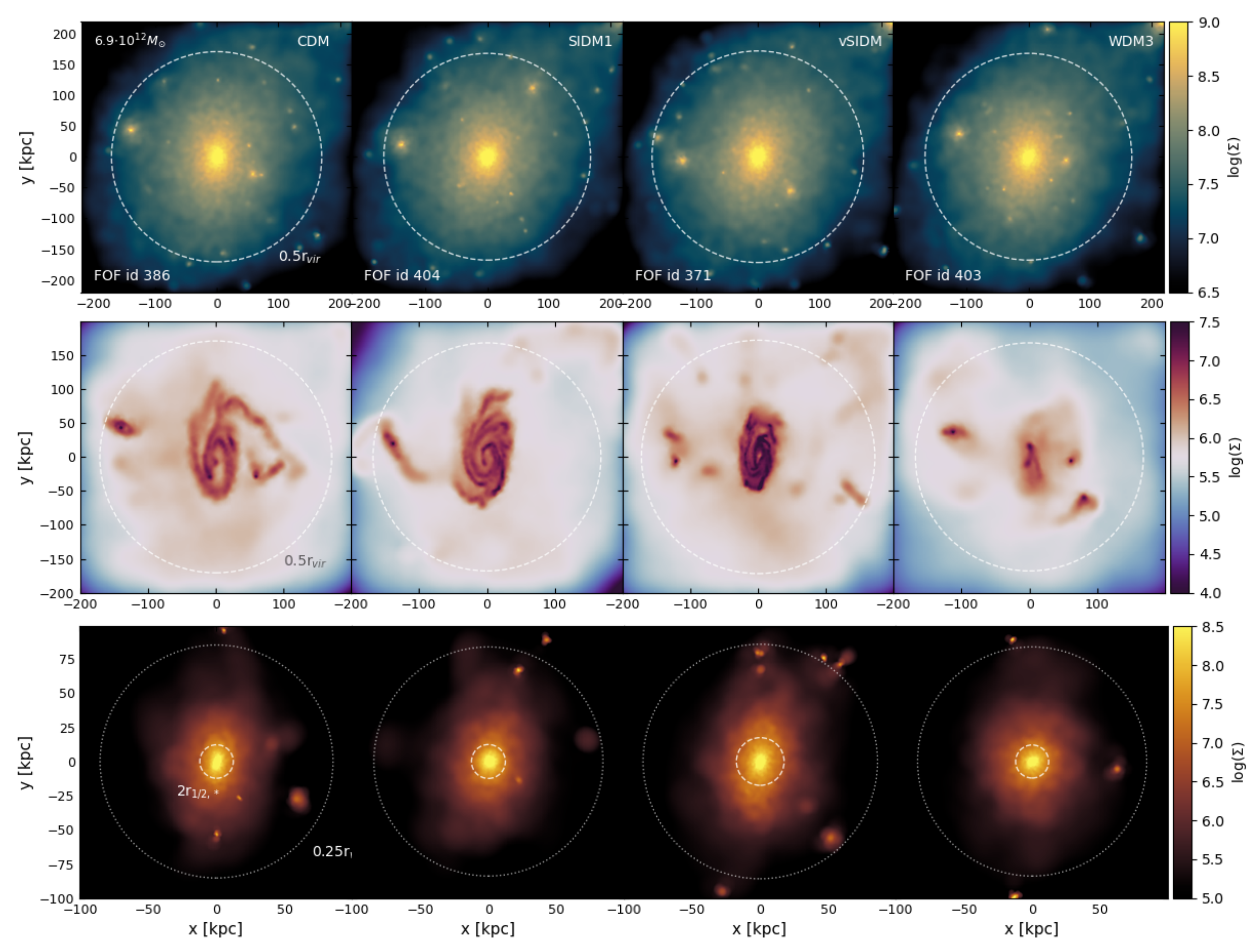}
   \caption{Visualisation of the dark matter (top), gas (middle), and stellar (bottom) projected mass distributions in a Milky-Way mass halo ($M_{\rm vir}=6.9\times10^{12}\,{\rm M}_{\odot}$ in CDM at $z=0$) from the 100/A runs. From left to right, CDM, SIDM1, vSIDM, and WDM3. We zoom-in to the central parts of the halo, $\sim0.5\,r_{\rm vir}$ in the top and middle panels, and $\sim0.25\,r_{\rm vir}$ in the stellar distribution. In the latter, we also mark the distance corresponding to twice the stellar half-mass radius. One can observe a decrease in the number of substructures in WDM3 as well as a clear difference in the gas dynamics and the shape of the stellar distribution. 
              \label{fig:visual1}%
    }
\end{figure*}
\begin{figure*}
    \centering
    \includegraphics[width=\textwidth]{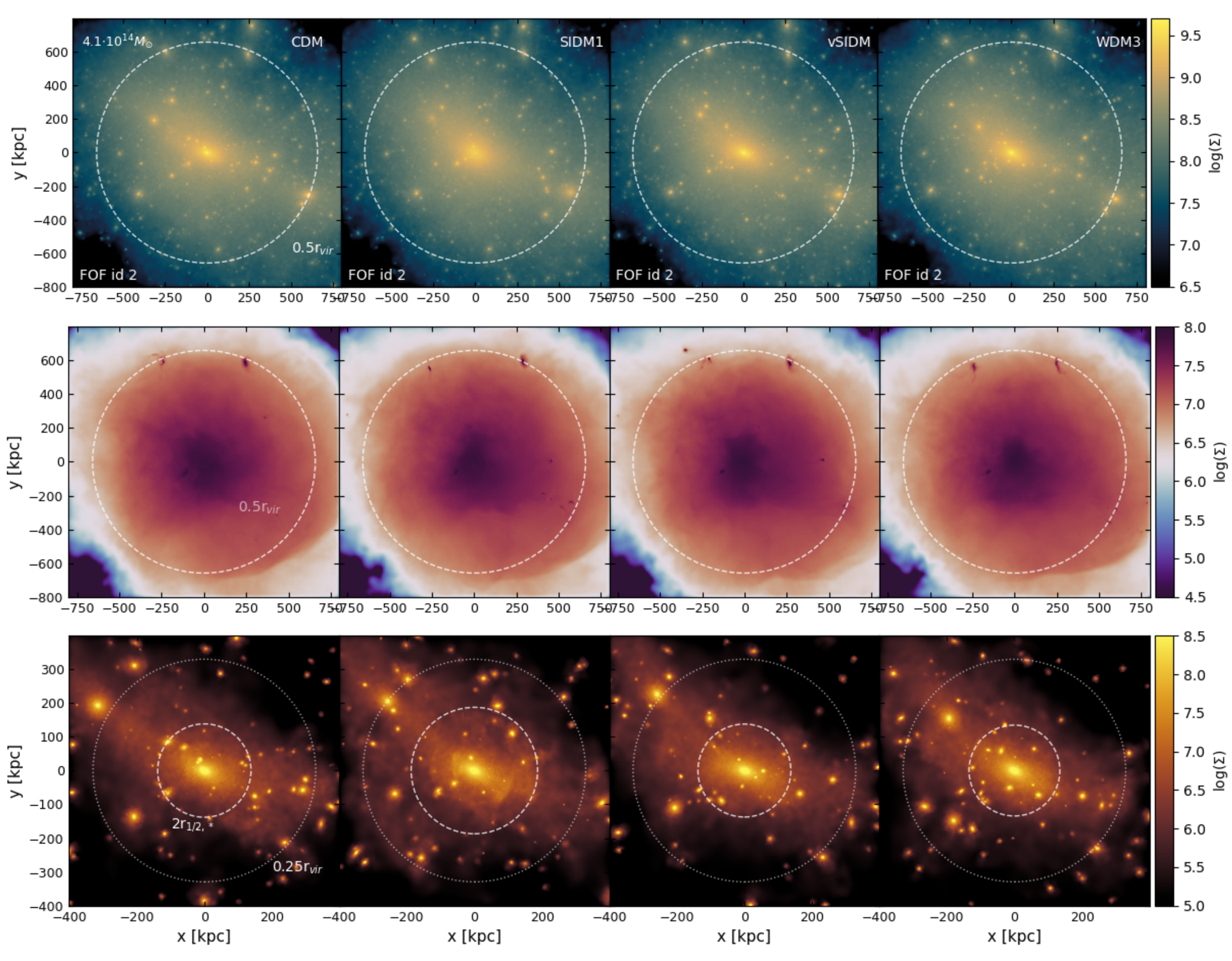}
   \caption{Same as Fig.~\ref{fig:visual1} but for a galaxy cluster of mass $M_{\rm vir}=4.1\times10^{14}\,{\rm M}_{\odot}$ in CDM at $z=0$. In this case, the SIDM1 model produces a more diffuse central halo (see the dark matter component in the top panel), also corresponding to a much larger effective radius (bottom panels).
              \label{fig:visual2}%
    }
\end{figure*}

\begin{figure*}
    \centering
    \includegraphics[width=\textwidth]{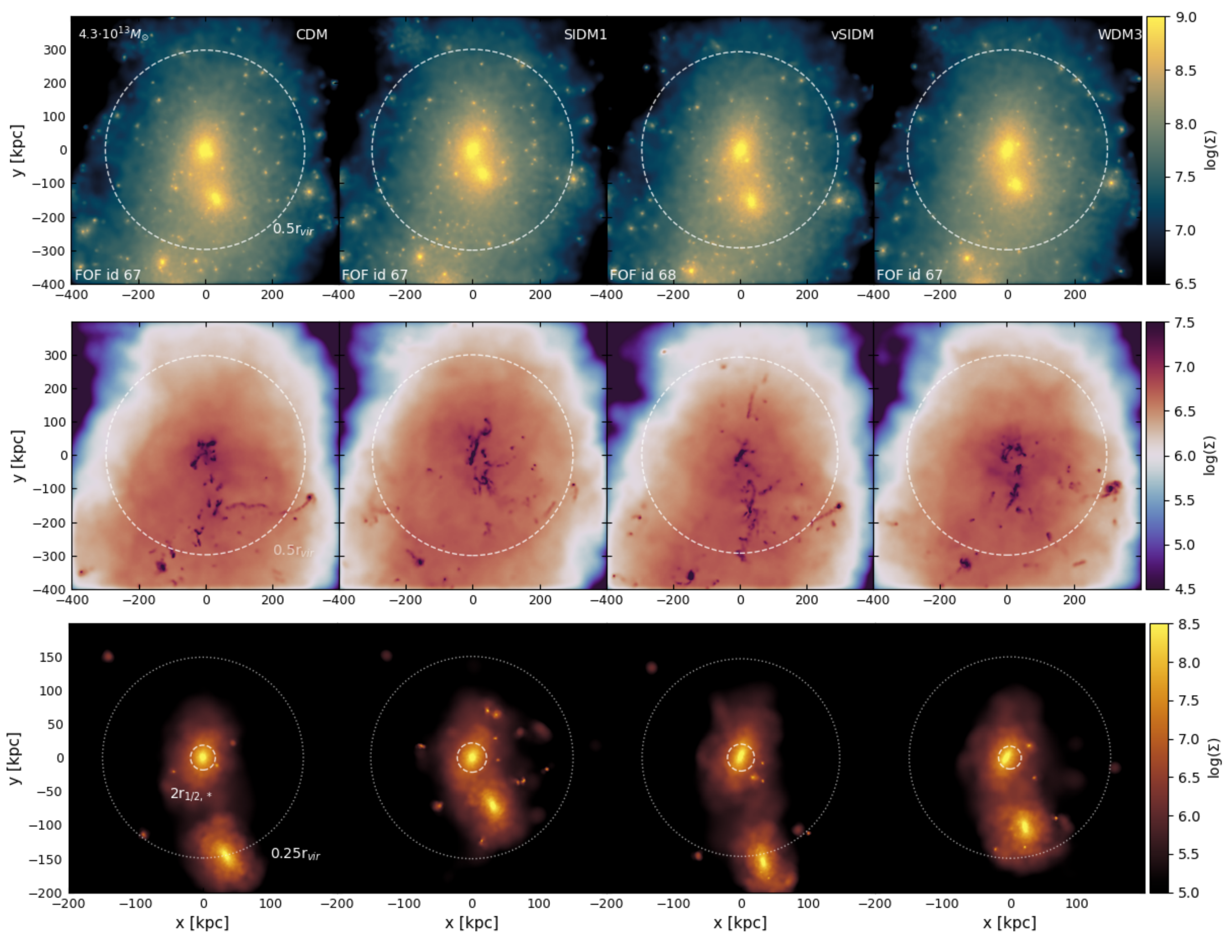}
   \caption{Same as Fig.~\ref{fig:visual1} but for a group-scale system with a CDM mass of $M_{\rm vir}=4.3\times10^{13}\,{\rm M}_{\odot}$ undergoing a major merger with a halo of mass $M_{\rm vir}=4.5\times10^{12}\,{\rm M}_{\odot}$. The merger happens around $z=0$ in all scenarios but the timescale differs.  \label{fig:visual3}
    }
\end{figure*}

\subsection{Self-interacting dark matter models} \label{subsec:sidm}

Self-interacting dark matter generally refers to a variety of models in which dark matter particles have both gravitational and non-gravitational interactions \citep[see][for a review]{tulin18}. In this scenario, a whole hidden phenomenology of dark interactions could exist, separated from the interaction with ordinary matter \citep{kaplinghat14}. Moreover, the self-interaction cross-section $\sigma$, which regulates the amount of collisions between particles, can be constant or inversely dependent on the relative velocity of particles. The interactions lead to an exchange of energy and momentum between particles, which does not take place in CDM or WDM, and induce modifications of the dark matter distribution within haloes. Over the course of cosmic time, high-density regions are partially eroded, leading to the formation of cores. However, the SIDM phenomenology is complex and, if the interactions are frequent and the core is quickly created, this can cause instability and the onset of gravothermal core collapse \citep{koda11,rocha13,turner21,yang21,zhong23}, leading to profiles that are cuspier than CDM instead of cores. 

Self-interactions leave a whole set of astrophysical signatures, spanning many orders of magnitudes from dwarf galaxies to galaxy clusters \citep[see][for a review]{adhikari22}. Observational constraints on the self-interaction cross-section can differ significantly, depending on the considered scale. For example, the elongation of haloes hosting elliptical galaxies yields $\sigma/m \leq 1\,{\rm  cm^{2}g^{-1}}$ \citep{peter13} and strong lensing arc statistics in galaxy clusters allows $\sigma/m_{\chi} < 0.1\,{\rm  cm^{2}g^{-1}} $ \citep{meneghetti01}. Given these disparities on different scales, a constant cross-section compatible with cluster scale limitations is unable to lower the central density of dwarf galaxies appreciably \citep{zavala13,fry15}. As a result, there has been a surge of interest in SIDM models with velocity-dependent cross-sections, where the strength of the interactions decreases with increasing relative velocity of the particles. The effective cross-section in dwarf galaxies can then be several orders of magnitude bigger than in cluster-sized haloes, bringing constraints on different scales into agreement \citep{zavala19,correa21,yang23}. 

Recent works are instead starting to model the interplay between self-interactions and baryonic physics, finding that previous predictions based on DMO simulations need to be updated. At the scale of (massive) galaxies, \citet{sameie18} used simulations of isolated galaxies to show that the cross-talk between SIDM and baryons produces a wide range of halo profiles, depending on how centrally concentrated the baryonic component is. \citet{rose23} found that the density profiles of haloes hosting MW galaxies are not cored in the presence of baryons and that SIDM haloes have a stronger response to baryons than CDM ones. \citet{despali19} ran full-hydrodynamic zoom-in simulations of nine haloes hosting massive galaxies, finding that SIDM haloes can be both cored and cuspy, depending on halo mass, morphological type, as well as the halo mass accretion history. Moreover, the same simulations show that CDM and SIDM halo shapes are similar when baryons are included \citep{despali22}, and thus $\sigma/m_{\chi} = 1\, {\rm cm^{2}g^{-1}}$ is not yet excluded, revising the constraints from \citet{peter13}. \citet{robertson18, robertson19,robertson21}, and \citet{shen22} studied the effect of SIDM on the density profiles of galaxy clusters and have concluded that the inclusion of baryons reduces the difference between CDM and SIDM predictions.

In this work, we consider two self-interacting models with constant and velocity-dependent cross-sections, shown in the right panel of Fig. \ref{fig:models}:

\begin{itemize}
    \item One is a classic elastic SIDM model with a constant cross-section $\sigma/m_{\chi}=1\,{\rm cm^{2}g^{-1}}$ (blue line) - SIDM1 hereafter. This model is already excluded at the scales of galaxy clusters, but it serves as an important comparison to previous works given that it has been used in most SIDM numerical experiments so far. Moreover, the constant cross-section more easily produces predictable effects and eases the interpretation of the other model. 
    \item The second is a model with elastic scattering where the interaction cross-section is velocity dependent (vSIDM hereafter). We chose the model by \citet{correa21} (green line) which is based on an empirical fit calibrated at the low-velocity end on observations of dwarf spheroidals in the Milky Way. At these scales, $\sigma/m_{\chi}=100\,{\rm cm^{2}g^{-1}}$, and thus we expect important effects, while the cross-section drops very rapidly at larger velocities (corresponding to high halo masses) in order to be consistent with observational limits at the scales of massive galaxies and clusters.
\end{itemize}

We used the self-interaction scheme implemented in \textsc{Arepo} by \citet{vogel12,vogel16,vogel19}. It corresponds to a particle physics model where the self-scattering between dark matter particles of mass $m_{\chi}$ is set by an attractive Yukawa potential mediated by a new gauge boson in the dark sector. A Monte Carlo approach is used to account for dark matter self-interactions where pairs of particles selected for collision are assigned new velocities with a randomly drawn direction. The time step is also chosen to be small enough to avoid multiple scatterings during one single step \citep[see][for more details]{vogel12}. The simulations start from the same ICs as the CDM ones, given that self-interactions do not affect the initial perturbations and become important during the late non-linear evolution of structures. 

\subsection{TNG galaxies in alternative dark matter} \label{subsec:tng}

We employ the TNG galaxy formation model described in detail in \citet{weinberger17,pillepich18}. 
We keep the fiducial TNG galaxy formation model entirely unchanged in all our runs despite the changes in the nature of dark matter. This way, we are able to characterise the effects of ADM models on galaxies and haloes and look at relative differences to their CDM counterparts. 

The TNG model has been extensively tested in large-scale simulations, namely the TNG50, TNG100, and TNG300 volumes \citep{springel18,pillepich18b,nelson18,marinacci18,naiman18,pillepich19,nelson19}, plus more recent projects including MillenniumTNG \citep{pakmor23}, TNG-Cluster \citep{nelson24} and the THESAN runs \citep{kannan22}. Moreover, it has been used in a number of smaller projects of zoom-in simulations both in cold \citep{lee24} and ADM models \citep{despali19,rose23}. It has been shown to roughly reproduce a broad range of galaxy properties and scaling relations across the galaxy population and as a function of the cosmic epoch. The model updates upon the previous Illustris model \citep{vogel14,Torrey14}, including a revised kinetic AGN feedback model for the low accretion state \citep{weinberger17} and an improved parametrisation of galactic winds \citep{pillepich18}. The model includes (i) a sub-resolution treatment of the interstellar medium (ISM) as a two-phase gas where cold clumps are embedded in a smooth, hot phase produced by supernova explosions \citep{springel03}; (ii) feedback from supernova explosions and stellar winds, in the form of kinetic and thermal energy; (iii) the production and evolution of nine elements (H, He, C, N, O, Ne, Mg, Si, and Fe), as well as the tracking of the overall gas metallicity and a model for Europium; (iv) density-, redshift-, metallicity-, and temperature-dependent cooling; and (v) super-massive black hole (SMBH) formation, growth and feedback in two different modes (quasar/thermal and kinetic/low-state).

In Figs. \ref{fig:visual1}, \ref{fig:visual2} and \ref{fig:visual3}, we show a visualisation of three systems to provide a qualitative picture of the effect of ADM at different scales and of the wealth of information available in the AIDA simulations. In each figure, the top panels show the projected density distribution of dark matter in the FP run of four dark matter scenarios: We focus on the central part of the halo ($\leq 0.5\, r_{\rm vir}$, represented by the white dashed circle) given that it is here that we expect the most significant differences between dark matter models. Similarly, the middle and bottom panels show the gas density within the same distance of the dark matter and stellar distribution in an even smaller area ($\leq 0.25\, r_{\rm vir}$). We have chosen these systems from a random selection within each mass bin in the 100/A runs and matched the CDM haloes to their counterparts in the other simulations.

The first galaxy in Fig. \ref{fig:visual1} lives in a halo of mass $M_{\rm vir}=6.7\times10^{12}\,{\rm M}_{\odot}$ at $z=0$ and has a total stellar mass of $M_{*}=1.3\times10^{11}\,{\rm M}_{\odot}$. Looking at the dark matter distribution (top panels), one can see that the overall structure is similar in all simulations, as well as the virial radii (and thus masses) of matched haloes. However, we observed $(i)$ the expected suppression of low-mass structures in WDM and $(ii)$ a more diffuse central peak in self-interacting models, especially SIDM1. We also observed clear differences in the small-scale distribution of the gas (middle panels) and its accretion and rotation pattern in the halo, which clearly differ in density and compactness -- or are disrupted in WDM3. Finally, the stellar distribution is more diffuse in the vSIDM model, as shown by the size of the stellar half-mass radius: the dashed circle, which represents $2r_{\rm eff,*}$. Figure~\ref{fig:visual2} shows a more massive system of cluster size: $M_{\rm vir}=4.1\times10^{14}\,{\rm M}_{\odot}$, $M_{*}=2\times10^{12}\,{\rm M}_{\odot}$ and $M_{\rm gas}=5.2\times10^{13}\,{\rm M}_{\odot}$. In this case, SIDM1 shows a particularly striking difference both in the dark matter central density and the size of the stellar distribution. In contrast, fewer differences are visible in the gas panels. Finally, in Fig.~\ref{fig:visual3}, we show a group mass halo undergoing a major merger. The merger affects the gas distribution similarly in all models, but the timescale is different, highlighting how the underlying dark matter model modifies the dynamics and timings of structure formation. These visualisations allow us to draw a few first qualitative considerations about the differences between the considered dark matter models -- even though these examples do not represent the full complexity of galaxy and halo properties, as well as environments that are present in the AIDA runs.


\section{Halo and stellar mass functions} \label{sec:massf}

\begin{figure*}
    \centering
   \includegraphics[width=\columnwidth]{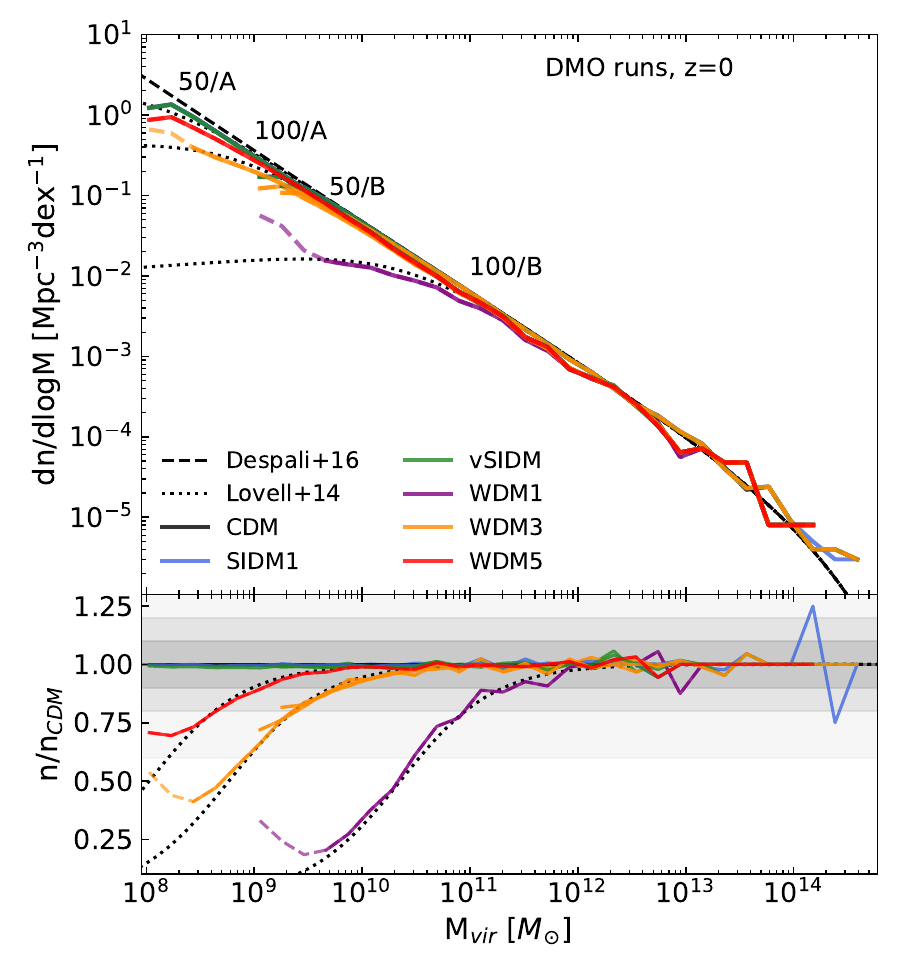}
   \includegraphics[width=\columnwidth]{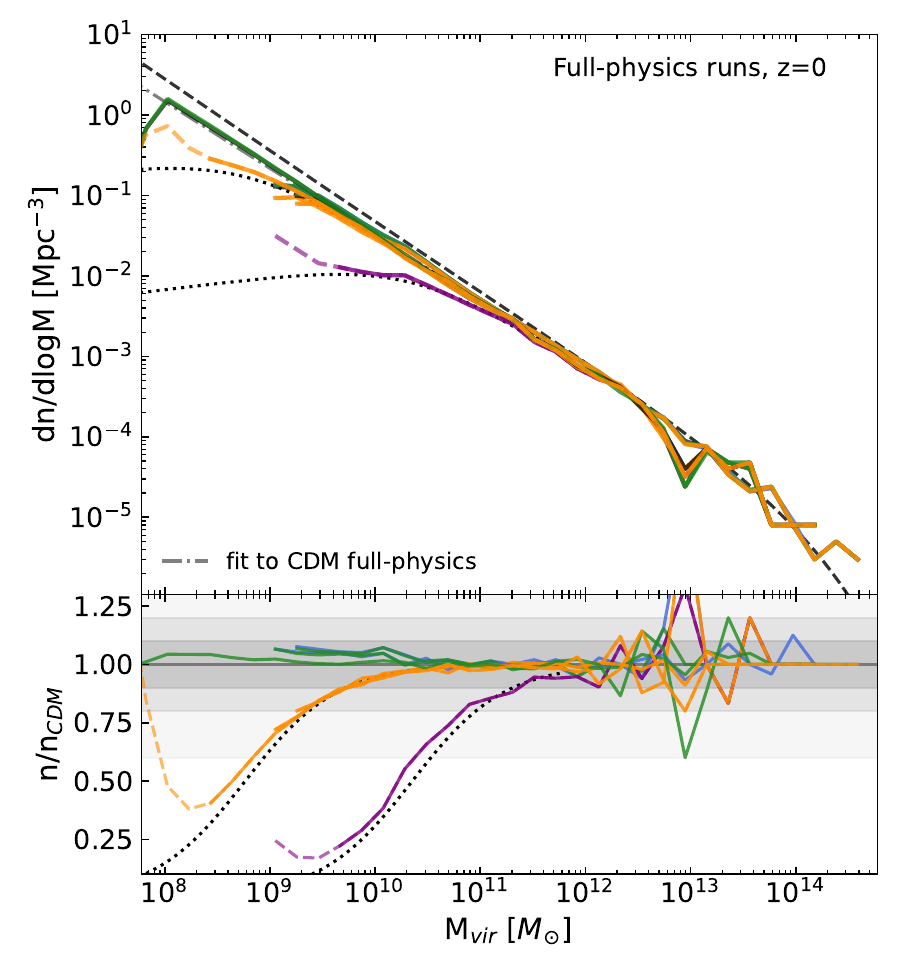}
   \caption{Halo mass function at $z=0$ in the DMO (left) and FP (right) runs of the four simulation sets listed in Table~\ref{table:1}. For each set, we only plot haloes with masses $M_{\rm vir}\geq100\,m_{\rm DM}$, corresponding to increasingly smaller low-mass limits as indicated by the labels in the top-left panel. The smaller panels show the ratio of each model to CDM. The dashed black line gives the theoretical prediction calculated by \citet{despali16} -- which agrees very well with the mass function measured in the DMO runs, while the dotted lines are calculated by applying to it the correction derived by \citet{lovell14}. In the right panel, we also plot the CDM DMO mass function (black dashed line) to highlight the suppression generated by baryons at the low-mass end, compared to the DMO case. For the WDM runs, the transition between dashed and solid lines marks the limiting mass $M_{\rm lim}$ below which the runs are affected by artificial fragmentation.}
              \label{fig:massf}%
\end{figure*}

\begin{figure*}
    \centering
   \includegraphics[width=\textwidth]{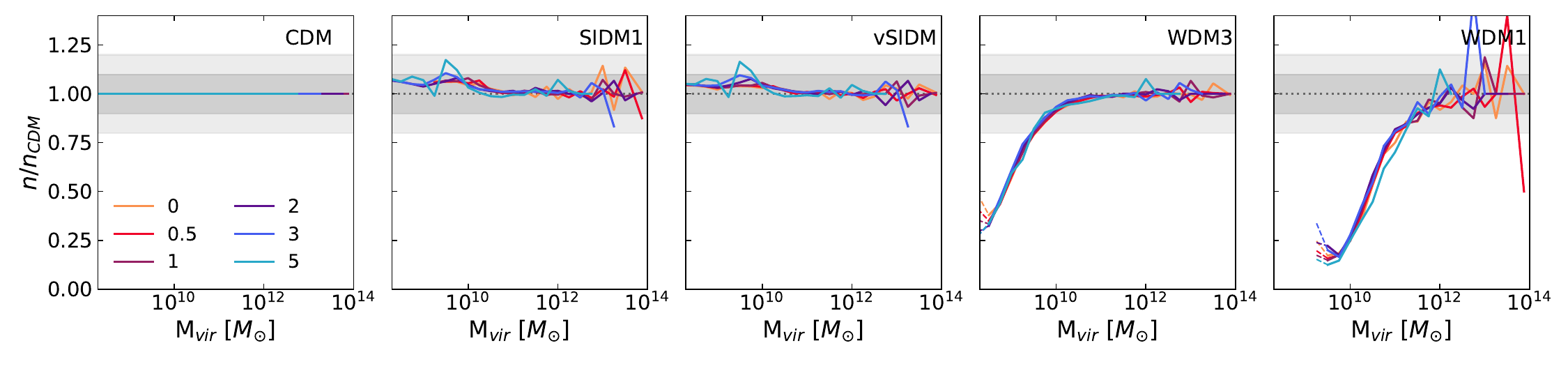}
   \caption{Redshift evolution of the effects of different dark matter models. In each panel, we show the ratios of each halo mass function to CDM at six different redshifts between $z=0$ (orange) and $z=5$ (light blue), calculated by averaging over all the available FP runs of each dark matter model. 
   }
              \label{fig:massf2}%
\end{figure*}

\begin{figure*}
    \centering
   \includegraphics[width=\textwidth]{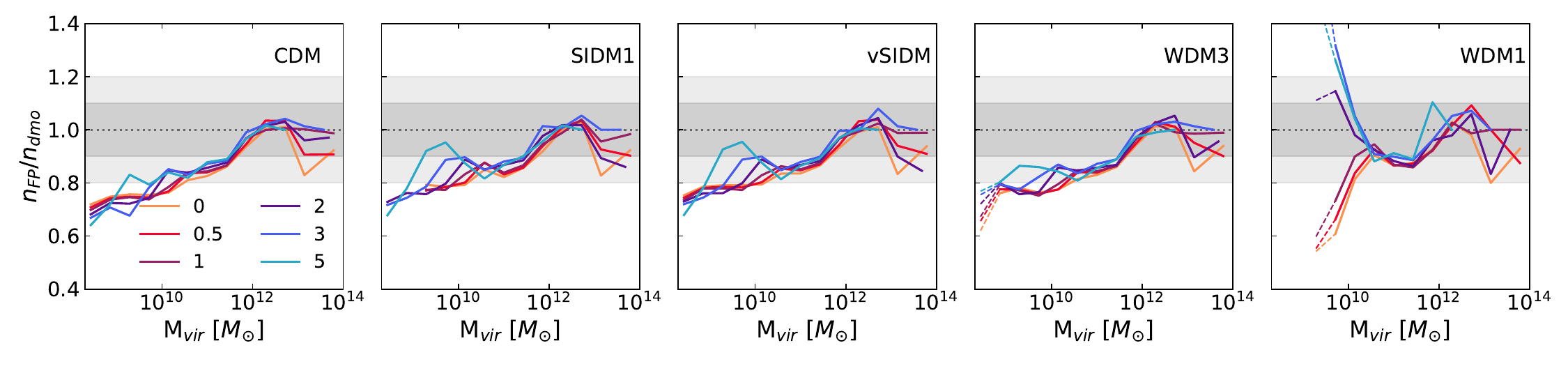}
   \caption{Baryonic effects on the halo mass function. In each panel, we compare the FP and DMO runs of each dark matter variation, plotting their ratios at six different redshifts from $z=0$ (orange) to $z=5$ (light blue), as in Fig. \ref{fig:massf2}. 
   }
              \label{fig:massf3}%
\end{figure*}

The low-mass end of the halo and subhalo mass functions is known to be affected by modifications to the dark matter particle mass in WDM models \citep{schneider12,lovell12,lovell14,lovell20} since the suppression of the small-scale power spectrum $P(k)$ directly alters the formation of objects at the low-mass end. In SIDM, \citet{zavala13} showed that only a high constant cross-section of $\sigma/m_{\chi}=10\,{\rm cm^{2}g^{-1}}$ modifies the number of subhaloes at the low-mass, and \citet{despali19} confirmed that the effect was negligible for $\sigma/m_{\chi}=1\,{\rm cm^{2}g^{-1}}$ using zoom-in simulations including the TNG model. Thus, we do not expect strong differences due to self-interactions alone. 

However, in AIDA the effects of ADM combine with those caused by baryonic physics: Stellar and AGN feedback redistribute matter within the haloes and produce a suppression in the halo and subhalo mass functions of an amount that depends on the details of the galaxy formation model \citep[typically up to $\sim20$\%, see][]{schaye15,vogel14,despali17b,lovell18,springel18}. For example, \citet{despali19} have shown that the WDM (7 keV sterile neutrino, equivalent to a 3 keV model) mass function of subhaloes hosted by massive galaxies is further depleted in the presence of baryons, using zoom-in runs of four haloes that included the EAGLE galaxy formation model. We now investigate both effects by computing the halo mass function in the DMO and FP runs and its evolution with redshift.

In the AIDA simulations, we can reliably measure the halo mass function over six orders of magnitude in halo mass and thus characterise the full halo population in all dark matter scenarios. We start by plotting the halo mass functions at redshift $z=0$ in Fig.~\ref{fig:massf} for halo masses in the range $10^{8}\leq M_{\rm vir}\leq 5\times10^{14}\,{\rm M}_{\odot}$. Here we choose the virial mass $M_{\rm vir}$ as our definition of halo mass. This is computed by \textsc{SUBFIND} for each FOF group in the halo catalogue and corresponds to the radius that encloses the virial overdensity $\Delta_{\rm vir}$ \citep{bryan98}. We show measurements for the dark (left) and FP (right) runs. In the bottom panels, we plot the ratio of each ADM mass function to CDM to highlight the mass range in which one can observe significant deviations. We also compared AIDA to theoretical predictions, namely, the \citet{despali16} virial mass function for CDM and the suppressed number density predicted for WDM by \citet{lovell14}
\begin{equation}
    \frac{n_{\rm WDM}}{n_{\rm CDM}}=\left(1+\gamma\frac{M_{\rm hm}}{M}\right)^{\beta}, \label{eq:lovell}
\end{equation}
\noindent where $\gamma=2.7$, $\beta=0.99$ and M$_{\rm hm}$ is the half-mode mass of the WDM model. As discussed in Sect.~\ref{subsec:wdm}, WDM simulations can be affected by artificial fragmentation below a limiting mass $M_{\rm lim}$, causing the identification of spurious structures. The transition to dashed lines in  Fig.~\ref{fig:massf} marks the regime where $M_{\rm vir}\leq M_{\rm lim}$. We found a good correspondence between $M_{\rm lim}$ and the upturn of the mass functions at low-mass haloes caused by artificial fragmentation. 

In the dark runs, we found excellent consistency between the theoretical predictions and our measurements since we observed the predicted suppression in WDM and instead did not find any significant difference in SIDM counts. In the right panel of Fig.~\ref{fig:massf}, the model by \citet{despali16} does not reproduce the CDM counts because stellar and AGN feedback suppresses the number density at low masses, as mentioned above. In this case, we fit for the slope of the mass function, finding $\alpha=-0.82$ (instead of $-0.9$) and used this as a baseline to calculate the predicted numbers in WDM instead of the prediction used in the left panel. The number of haloes with masses M$_{\rm vir}<10^{10}\,{\rm M}_{\odot}$ is slightly above CDM both in WDM and SIDM, consistent with the fact that also Eq.~(\ref{eq:lovell}) was derived for DMO runs.

Next, we look at the redshift evolution of the halo mass function and its dependence on ADM and baryonic effects. In each panel of Fig.~\ref{fig:massf2}, we show the ratio of the halo mass function in one scenario to the reference CDM measurement, considering six snapshots at redshifts between $z=0$ and $z=5$. At all considered times, the ADM effects are remarkably similar to each other and to the $z=0$ case. In SIDM, the 5-10\% excess in the number density of haloes with M$_{\rm vir}\leq5\times10^{10}\,{\rm M}_{\odot}$ persists at all redshifts. Our WDM measurements are consistent with those by \citet{lovell24}, who recently used the COCO simulation to analyse the halo mass function up to redshift $z=15$ in a WDM model with particle mass $m_{\rm WDM}=3.3\,{\rm keV}$ -- thus similar to our WDM3 run. They showed that the suppression in WDM stays consistent for most redshifts, with some differences appearing only at $z>10$. 

We then calculated the effect of baryons on the halo mass function. In Fig. \ref{fig:massf3}, we show the ratios between the counts in the FP and the corresponding DMO runs in each model and again consider six redshifts between $z=0$ (orange) and $z=5$ (light blue). Interestingly, baryonic effects are quite similar in all dark matter scenarios. The only exception is the WDM1 run, where the number of low-mass haloes is higher in the presence of baryons at $z\geq2$, despite the mass function being still depleted compared to CDM. However, this could be a non-trivial consequence of artificial fragmentation and its interplay with the presence of baryons. The similarity between dark matter scenarios is a remarkable result, considering that the TNG galaxy formation model has been calibrated in the standard CDM, and we have not modified it in any way to absorb ADM effects. This happens despite the striking differences in the number density of WDM haloes and the structural properties of haloes in both alternative scenarios (such as the density profiles discussed later in Sect.~\ref{sec:prof}). 

\begin{figure}
    \centering
   \includegraphics[width=\columnwidth]{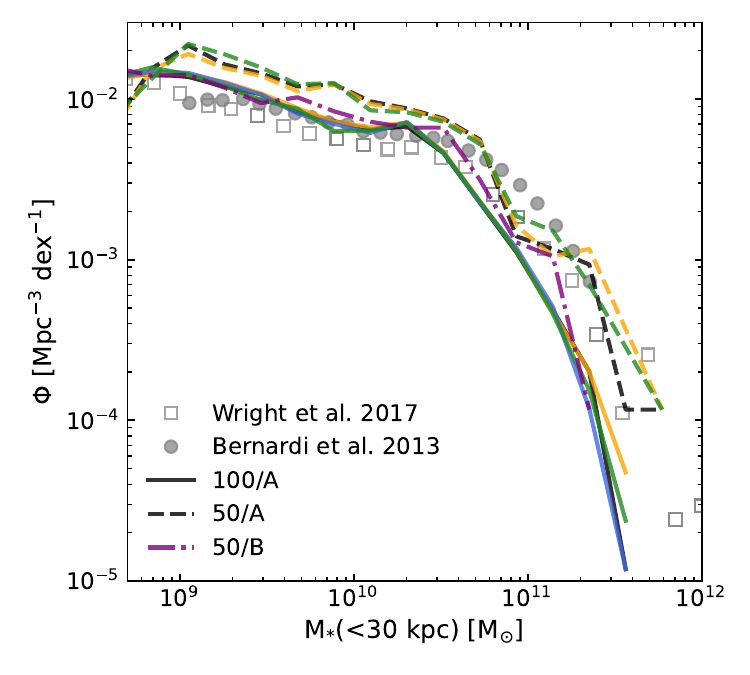}
   \caption{Stellar mass function at $z=0$. In each run, we consider the stellar mass measured within $r=30$ kpc from the halo centre and qualitatively compare it to observations from SDSS \citep{bernardi13} and GAMA \citep{wright17}. Solid and dashed lines represent the 100/A and 50/A runs, while we show the 50/B measurements (dot-dashed) only for the WDM1 scenarios. The colour scheme of the different dark matter models is the same as in Fig.~\ref{fig:massf}.
   }
              \label{fig:massf_st}%
\end{figure}

Finally, we analyse the stellar mass function at $z=0$. This is one of the quantities that has been considered in the calibration of the TNG model, and thus, we want to check if it is affected by dark matter physics. In Fig.~\ref{fig:massf_st}, we qualitatively compare the simulated results to observations from SDSS \citep{bernardi13} and GAMA \citep{wright17}. We do not observe systematic differences between different dark matter scenarios at fixed resolution, consistent with the fact that all halo mass functions predict the same number of haloes at $M_{\rm vir}\geq 5\times10^{11}\,{\rm M}_{\odot}$ (see the right panel in Fig.~\ref{fig:massf}). This means that, despite the differences at the low-mass end, the models considered here cannot be distinguished in terms of the number of observed galaxies, even in the most extreme cases. The difference due to resolution is consistent with the previous analyses by \citet{pillepich18} and \citet{pakmor23}. Moreover, comparing our results to the analysis by \cite{pillepich18}, we can conclude that variations in the stellar mass function due to ADM models are much smaller than those produced by a modification of the baryonic treatment. In Sect.~\ref{sec:galaxies}, we return to considering other galaxy properties that have been used in the model calibration and discuss the scaling relations of galaxies in more detail.

\section{The structure of dark matter haloes} \label{sec:prof}

\begin{figure*}
   \includegraphics[width=\linewidth,center]{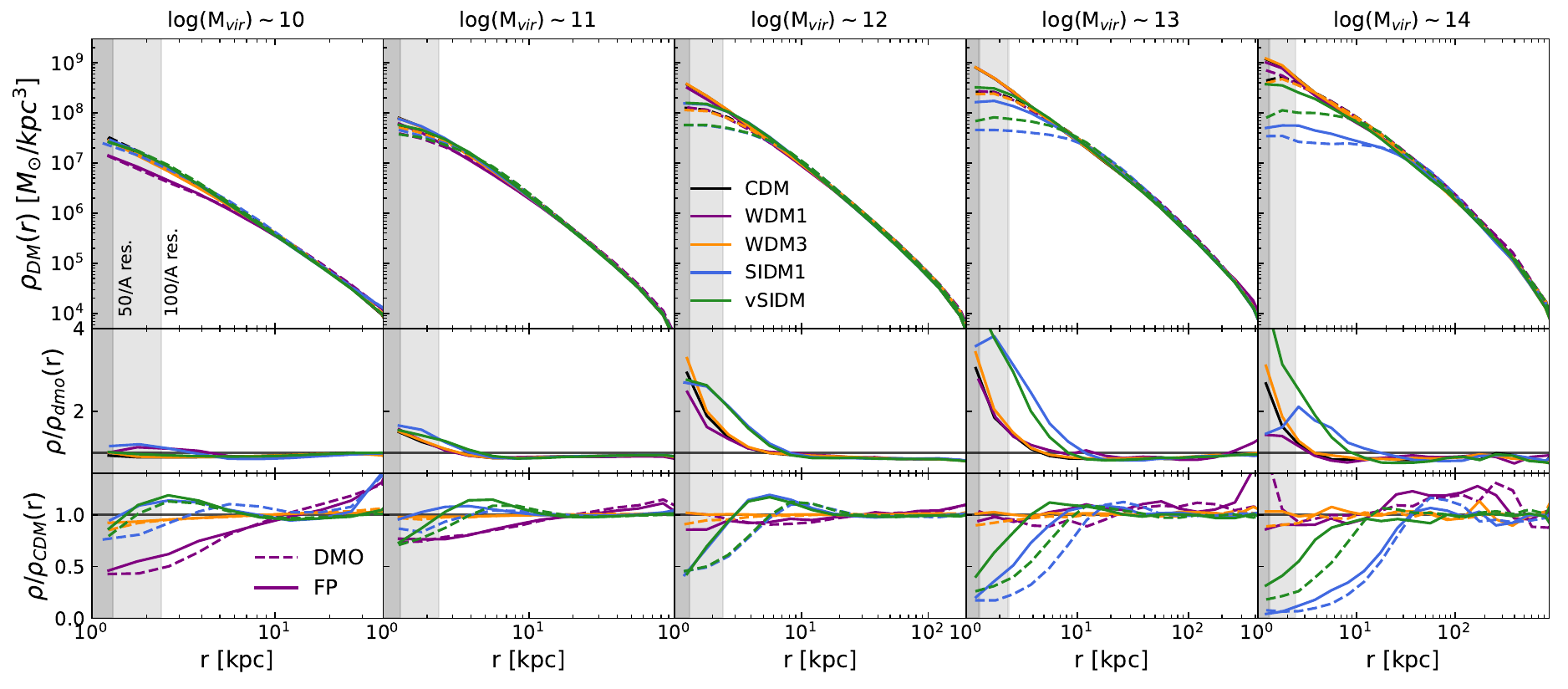}

   \caption{Dark matter density profiles at $z=0$ when considering five halo mass bins (left to right). We calculated the mean dark matter profile in bins of $\Delta \log(M_{\rm vir})=0.2$ dex around the mean value, both in the FP (solid) and dark (dashed) runs. The profiles were calculated in logarithmically spaced spherical shells from 1 kpc to the halo virial radius. In order to capture both good statistics at high masses and the highest resolution, we average profiles from both the 50/A and 100/A runs for all models -- except WDM1, where 50/B is employed. The small panels show $(i)$ the ratio of each mean profile to its dark counterpart and $(ii)$ to the CDM version. We observe an opposite trend in mass in the WDM and SIDM models. The latter creates larger density cores at high masses and cuspy profiles at $M_{\rm vir}\leq 10^{12}\,{\rm M}_{\odot}$, while the former lowers the central density at low masses (halo masses below the half-mode mass $M_{\rm hm}$). In all panels, the grey bands indicate 2.3$\epsilon_{\rm DM}$ for both runs, which is often used as the minimum distance for reliable measurements.} 
              \label{fig:prof1}%
\end{figure*}

\begin{figure*}
    \includegraphics[width=\linewidth,center]{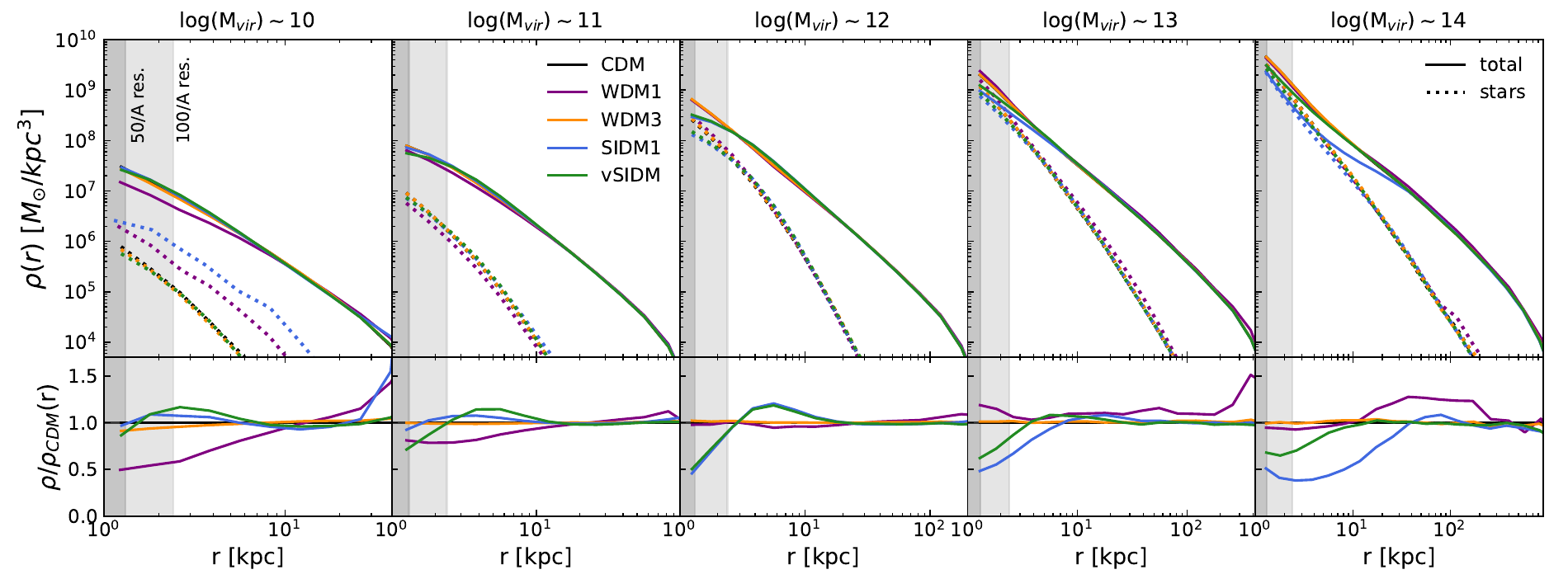}
   \caption{Total and stellar mean density profiles in the inner parts of the haloes at $z=0$, represented by solid and dashed lines. The mass bins are the same as in Fig.~\ref{fig:prof1}, and the bottom panels highlight the effect of alternative models on the total density profiles. Similarly to the dark matter profiles, the total density is affected by core formation in SIDM at the high-mass end, and in WDM at the low-mass end, but the core size and depth are reduced.} 
              \label{fig:prof2}%
\end{figure*}

Warm and self-interacting models alter how dark matter particles interact, thus impacting the internal structure of haloes and the galaxies they host. In this section, we look at the inner parts of haloes and quantify the effects of WDM and SIDM on density profiles of dark matter (Fig. \ref{fig:prof1}) and baryons (Fig. \ref{fig:prof2}), as well as on the halo concentration-mass relation.

\begin{figure*}
    \includegraphics[width=0.48\linewidth]{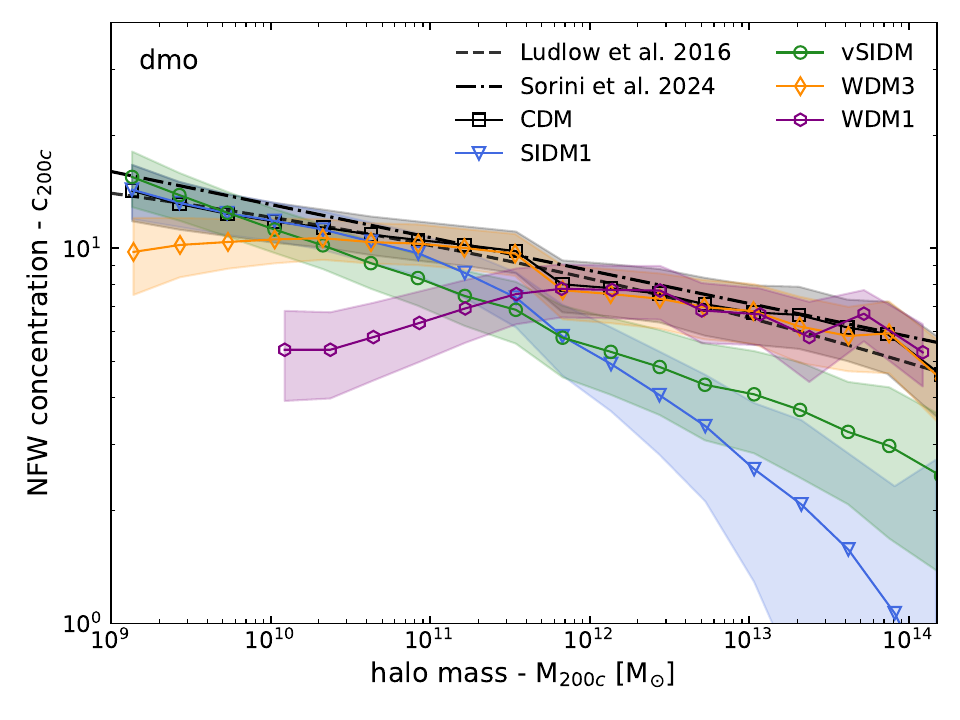}
    \includegraphics[width=0.48\linewidth]{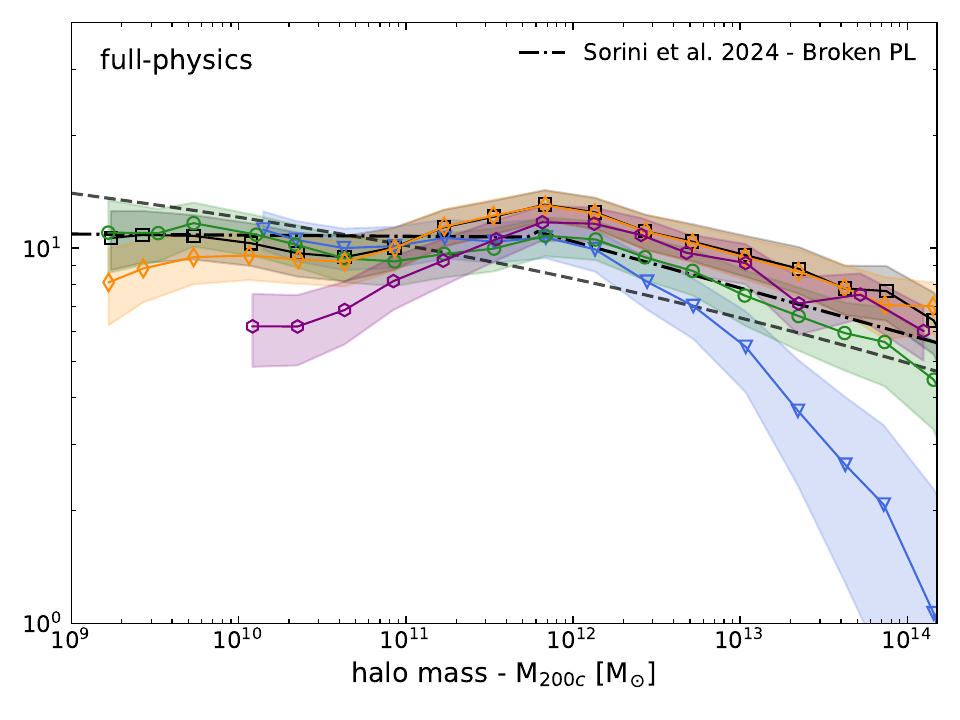}
   \caption{Concentration-mass relation at $z=0$ for haloes in the DMO (left) and FP (right) runs. We combine measurements from the 50/A and 100/A boxes, using the latter only for haloes with a mass $M_{\rm 200c}\geq10^{11}\,{\rm M}_{\odot}$. The coloured lines and symbols show the mean concentration as a function of mass, together with the 1-$\sigma$ uncertainty represented by the bands of the corresponding colour. The dashed and dashed-dotted black lines show the previous work by \citet{ludlow16} and \citet{sorini24} on CDM simulations instead. The latter used the IllustrisTNG and MillenniumTNG, which also employ the TNG galaxy formation model, and provided a broken-power-law fit for the FP runs, which we show here in the right panel.} 
              \label{fig:prof3}%
\end{figure*}

\subsection{Halo density profiles}

The AIDA boxes allow us to explore many orders of magnitude in halo mass as well as to measure the density distribution inside haloes down to kilo-parsec scales. We only computed density profiles of haloes resolved by at least 500 dark matter particles and thus, the 51.7 Mpc boxes are essential to reliably measure density profiles of haloes with $M_{\rm vir}\leq10^{10}\,{\rm M}_{\odot}$, thanks to the higher resolution and smaller particle mass. At the same time, the larger boxes provide us with the best statistical sample, especially at the high-mass end, $M_{\rm vir}\geq10^{13}\,{\rm M}_{\odot}$. We measured the density profiles of each halo by binning particles in 30 logarithmically spaced bins in the range between 0.4 kpc and the virial radius of the halo. The density within each radial bin is then computed as the ratio of the total mass of dark matter particles falling within said bin, and the volume enclosed between the spherical shells defined by the boundaries of the bin. To compute the total density profiles, we followed the same procedure but consider the mass in stars, gas and SMBHs in addition to dark matter. 

In the top panels of Fig.~\ref{fig:prof1}, we show the $z=0$ dark matter profiles for five representative bins in halo mass, from $10^{10}$ to $10^{14}\,{\rm M}_{\odot}$. In each bin, we selected haloes in a narrow range $\Delta\log M_{\rm vir}=0.2$ dex around each central value and then calculated the mean density profile of each model in logarithmically spaced bins out to the halo virial radius, both in the DMO (dashed lines) and FP (solid lines) runs. For each mass bin, we also use the sub-panels to highlight the effect of baryons (middle) and dark matter physics (bottom) on the dark matter distribution. Given that baryons and dark matter coexist at the centre of haloes, we expect that the variations in the dark matter distribution will also affect the baryonic content and vice versa. The general effect of baryons is to enhance the dark matter density in the central region, with the strongest effect being at $M_{\rm vir}\geq10^{13}\,{\rm M}_{\odot}$ (see the middle sub-panels) where the star formation and baryon fraction in the inner region are the highest. However, the magnitude of the baryonic enhancement depends on the dark matter physics, highlighting that the two effects are interconnected.

Both warm and self-interacting dark matter can create density cores, but the mechanisms that leads to the core formation and the mass ranges are different. The high velocities of WDM particles can naturally lower the concentration of structures at the low-mass end of the mass function, starting from haloes with a mass approximately one hundred times the model half-mode mass $M_{\rm hm}$ \citep{ludlow16}. Based on the half-mode mass values mentioned in Sect.~\ref{subsec:wdm}, we can thus expect cores appearing at $M_{\rm vir}\leq 2\times10^{12}\,{\rm M}_{\odot}$ in WDM1 and $M_{\rm vir}\leq4.8\times10^{10}\,{\rm M}_{\odot}$ in WDM3. However, this is the mass scale at which the halo concentrations start departing from the CDM values, but a significant effect is only present one order of magnitude below. 

In Fig.~\ref{fig:prof1}, we indeed observe a strong deviation of WDM1 profiles (purple lines) at $M_{\rm vir}\leq10^{11}\,{\rm M}_{\odot}$ where an inner dark matter core is created both in the DMO (dashed) and FP (solid) runs, while the model behaves similarly to CDM at higher masses. WDM3 profiles are instead more similar to the cold case, but a departure from CDM in the inner parts of the halo can be observed in the lowest mass bin. A different mechanism is at play in SIDM, where the density cores are created over time when self-interactions are triggered in the high-density central regions \citep{sameie18,despali20}. Both SIDM1 and vSIDM cores are thus larger -- and deeper -- for high-mass haloes and decrease at lower masses. The strongest effects are seen in the constant cross-section SIDM1 model (blue lines): In the dark runs, cores reach a size of $\sim10$ kpc for $10^{13}\,{\rm M}_{\odot}$ haloes and $\sim30$ kpc for $10^{14}\,{\rm M}_{\odot}$. When interactions are velocity-dependent, the core sizes are smaller in these two mass bins, while they are identical at lower masses. In the FP run (solid lines), the presence of a massive galaxy helps to reduce the core size (although the core is still present), while at $M_{\rm vir}\leq10^{12}\,{\rm M}_{\odot}$ we observe a $5$ to $10$\% increase of the central density compared to CDM, followed by a small central core which is, however, below our resolution limit and can thus be partly artificial. 

It is well known that two-body relaxation creates artificial cores at the centre of CDM haloes at a scale close to the softening length of the simulation and this has led to the adoption of $r\geq2.3\,\epsilon_{\rm DM}$ as reliable distances. The same effect is also present in the other dark matter models, as demonstrated by the fact that the $\rho_{\rm WDM}/\rho_{\rm CDM}$ ratios (bottom sub-panels) are close to unity up to the centre of the halo at high masses, where WDM effects do not alter the density distribution \citep{power03,hopkins18}. The SIDM central core in the mass bins $M_{\rm vir}\leq10^{11-12}\,{\rm M}_{\odot}$ is instead much deeper and thus arises due to a combination of numerical effects and self-interactions. 

The formation of a cuspy profile in the mass range of galaxies is consistent with the findings of previous works. For example, \citet{despali19} reported similar trends in the profiles (cored at the high-mass end but cuspy at intermediate masses) in zoom-in simulations of massive galaxies run with the SIDM1 and the TNG model. \citet{correa24} and \citet{rose23} also measured cuspy profiles in SIDM for Milky-Way mass haloes around $10^{12}\,{\rm M}_{\odot}$, which can be steeper than CDM. Notably, the effect is similar despite differences in the numerical implementation (SWIFT versus AREPO) and galaxy formation model (EAGLE versus TNG). \citet{rose23} concluded that, at the scale of Milky Way galaxies, baryons change the thermal structure of the central region of the halo to a greater extent than the SIDM scattering, and these changes cause SIDM to create cuspier central densities rather than cores because the SIDM
scatterings remove the thermal support by transferring heat away from the centre of the galaxy. The EAGLE50 and BAHAMAS-SIDM simulations that include the Eagle galaxy formation model \citep{robertson19,robertson21} also produce large cores, $\sim30$ kpc for $10^{14}\,{\rm M}_{\odot}$. However, in our simulations, we find a stronger effect of baryons on the dark matter profiles (middle panels of Fig.~\ref{fig:prof1}) at $M_{\rm vir}=10^{14}\,{\rm M}_{\odot}$ compared to their work. \citet{robertson19} does not observe the steepening of the profile at the scale of massive galaxies and the density profiles at $10^{12-13}\,{\rm M}_{\odot}$ for the model with constant cross-section $\sigma=1\,{\rm cm^{2}g^{-1}}$ remain shallower in SIDM than CDM. This is the halo mass range where star formation is the most efficient, baryons dominate the inner density profile, and the dark matter nature plays a secondary role. Thus, it could be heavily affected by the galaxy formation model.

In Fig.~\ref{fig:prof2} we switch to the total and stellar density profiles in the full physics runs and the same mass bins of Fig.~\ref{fig:prof1}. Even if stars are a significant or dominant mass component at the centre (see the dotted lines), the profiles are still affected by the different clustering of dark matter in the halo formation and, indeed, we observe that the total density behaves similarly to the dark matter distribution from Fig.~\ref{fig:prof1}, even if the effects are weaker. In addition to the maximum reduction that we measure, it is also important to evaluate at which dynamical scale within the halo the core becomes important: While SIDM cores are larger in massive haloes in physical units, the inner 10-20 kpc corresponds to less than 0.05\% of the virial radius, the WDM core created in low-mass haloes corresponds to 15-30\% of $R_{\rm vir}$. These statistical differences in the density profiles impact astrophysical observables related to the gravitational effects, such as gravitational lensing.  At the high-mass end between clusters and massive galaxies, the density cores in the total density profile modify the distribution of Einstein radii and lead to smaller angular separations between the lensed images. Low-mass haloes are instead not dense enough to lens background sources on their own, but they can be detected if they produce additional distortions on lensed arcs \citep{vegetti09}. In WDM, the low concentration of low-mass haloes reduces the number of detectable systems as discussed in \citet{gilman19} and \citet{amorisco21}. The AIDA simulations are thus ideal for comparisons with observational results at multiple scales through analytical predictions and the creation of realistic mock observations.

\begin{figure*}
    \centering
   \includegraphics[width=0.95\linewidth]{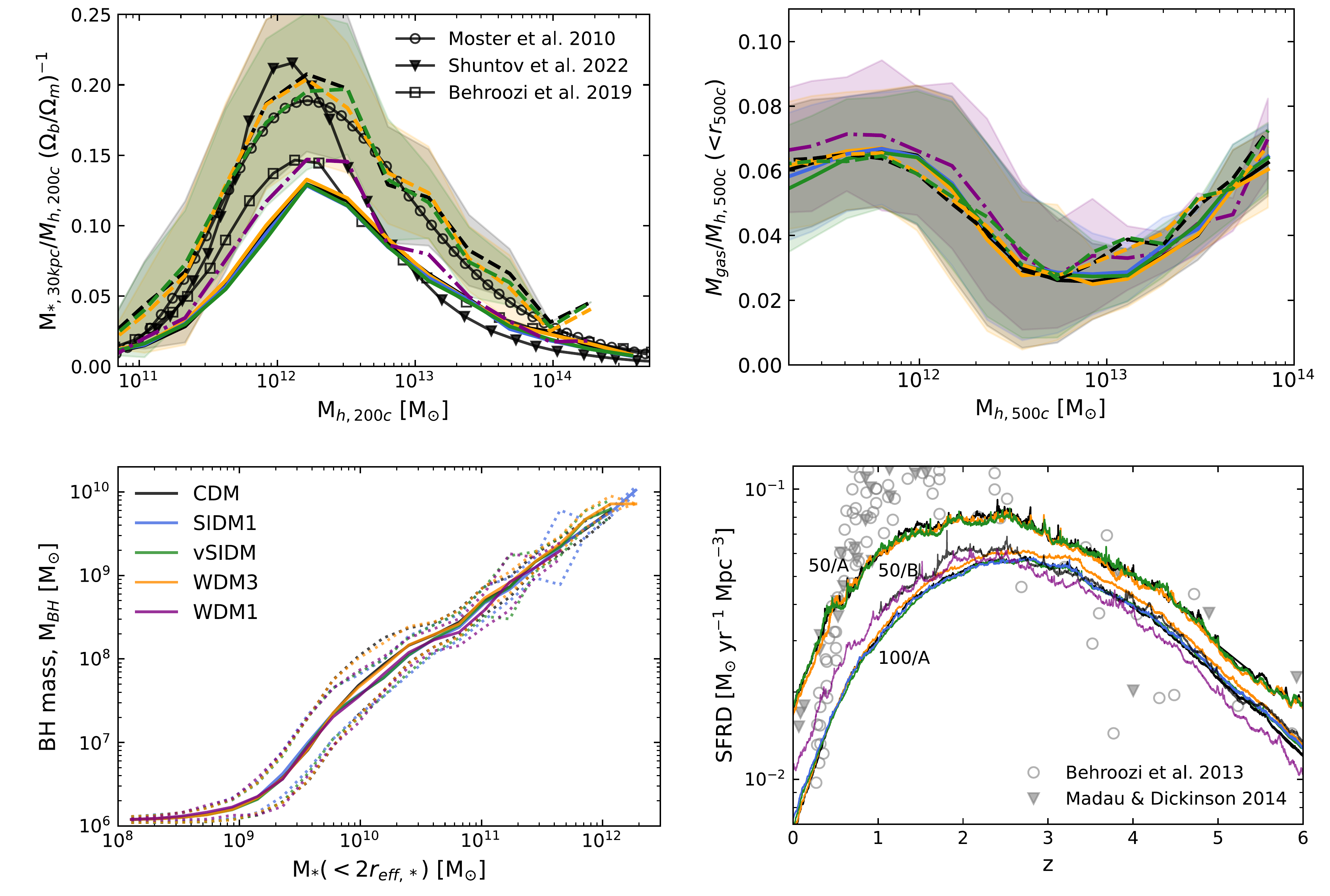}

   \caption{Quantitative tests of the TNG galaxy formation model in all dark matter scenarios at $z=0$. We focus on some of the quantities that have been used in the design of the TNG model \citep[see][]{pillepich18} together with the stellar mass function (Fig. \ref{fig:massf_st}) and the galaxy sizes (Fig. \ref{fig:gal_size}). The colour-correspondence to the dark matter model is the same in all panels. \emph{Top left}: Stellar-mass halo-mass relation. Here, $M_{*}$ is measured within 30 kpc from the centre. The black dotted curves show empirical estimates based on CDM and the model by \citet{moster10}, \citet{behroozi19} and \citet{shuntov22}, while the coloured curves show the results from our simulations, represented by running mean and standard deviations for each dark matter scenario. \emph{Top right}: Gas fraction of haloes within $r_{500c}$. \emph{Bottom left}: Central SMBH mass as a function of the stellar mass (here calculated within twice the stellar half-mass radius $r_{*}$) for all subhaloes at $z=0$. The dotted lines mark the 1$\sigma$ region. \emph{Bottom right}: Cosmic SFRD as a function of redshift compared qualitatively to observational measurements by \citet{behroozi13,madau14}.}
              \label{fig:gal_prop}%
\end{figure*}

\subsection{The concentration-mass relation at $z=0$}

In light of the results of the previous section, here we discuss the halo concentration-mass relation that summarised the properties of density profiles over the whole range of halo masses. We fit the dark matter density with an NFW profile \citep[Navarro-Frenk-White, see][]{navarro96} and calculated the concentration as $c_{200c}=R_{200c}/r_{s}$, where $r_{s}$ is the scale radius of the NFW fit and $R_{200c}$ is the radius that encloses an overdensity equal to $200\,\rho_{c}$. In Fig. \ref{fig:prof3}, we plot the concentration-mass relation in the considered five dark matter scenarios. We measure the concentration of the dark matter haloes both in the dark (left panel) and FP (right panel) runs to evaluate the response of the halo to the presence of baryons. The fit was performed in logarithmic scale, and we considered only the radial bins in the range between 2.3$\epsilon_{\rm DM}$ and $R_{200c}$. For each halo, we obtained the two parameters $\rho_{s}$ and $r_{s}$ and calculated the halo concentration as $c_{200c}$. We chose this as opposed to the virial radius $R_{\rm vir}$ to compare with previous results \citep{ludlow16,sorini24}, with which we found good consistency in CDM. In WDM and SIDM, the core formation causes a decrease in concentration at opposite ends of the halo mass range. We observe the turnover at $M_{\rm 200c}\simeq10^{12}\,{\rm M}_{\odot}$ for WDM1 and $M_{\rm 200c}\simeq10^{11}\,{\rm M}_{\odot}$ for WDM3, corresponding to $\sim$100$\, M_{\rm hm}$, as discussed above. Given that the SIDM1 model produces the largest cores, it also shows the largest departure from the CDM concentration-mass relation towards at high masses. Instead, the vSIDM model is less extreme at cluster masses but starts departing from CDM at lower masses compared to SIDM1. 

In the right panel, we show the concentration-mass relation measured in our FP runs and calculated by fitting the dark matter density profile of haloes. The concentration increases compared to the dark matter runs, peaking at $M_{200c}\sim10^{12}\,{\rm M}_{\odot}$ and reaching a plateau at low masses. This is due to the adiabatic contraction of infalling gas, which promotes star formation, increasing the central density and facilitating further dark matter collapse into the central regions of the halo. We compare our measurements to the broken power-law best fit (dot-dashed line) by \citet{sorini24} using the IllustrisTNG and MillenniumTNG runs, which employ the same galaxy formation model used here. Due to the presence of the central baryonic component, the effects of ADM models are reduced but maintain the same trend: The WDM suppression starts at a $lower$ mass compared to the dark runs, while the SIDM one does at $higher$ mass. 

In all cases, we fit an NFW profile to the density profiles. In the presence of large cores, we already know that this model is a priori not a good description of the data, which may be better fit by a more flexible Einasto profile \citep{einasto65}. Nevertheless, the NFW concentrations are a good indicator of the departure of the profiles from the standard CDM model. In a follow-up paper, we will model the redshift evolution of density profiles and of the concentration-mass relation, and provide a better fitting function for ADM models.

\section{Galaxy properties in alternative dark matter models} \label{sec:galaxies}

\begin{figure*}
    \centering
   \includegraphics[width=0.9\linewidth]{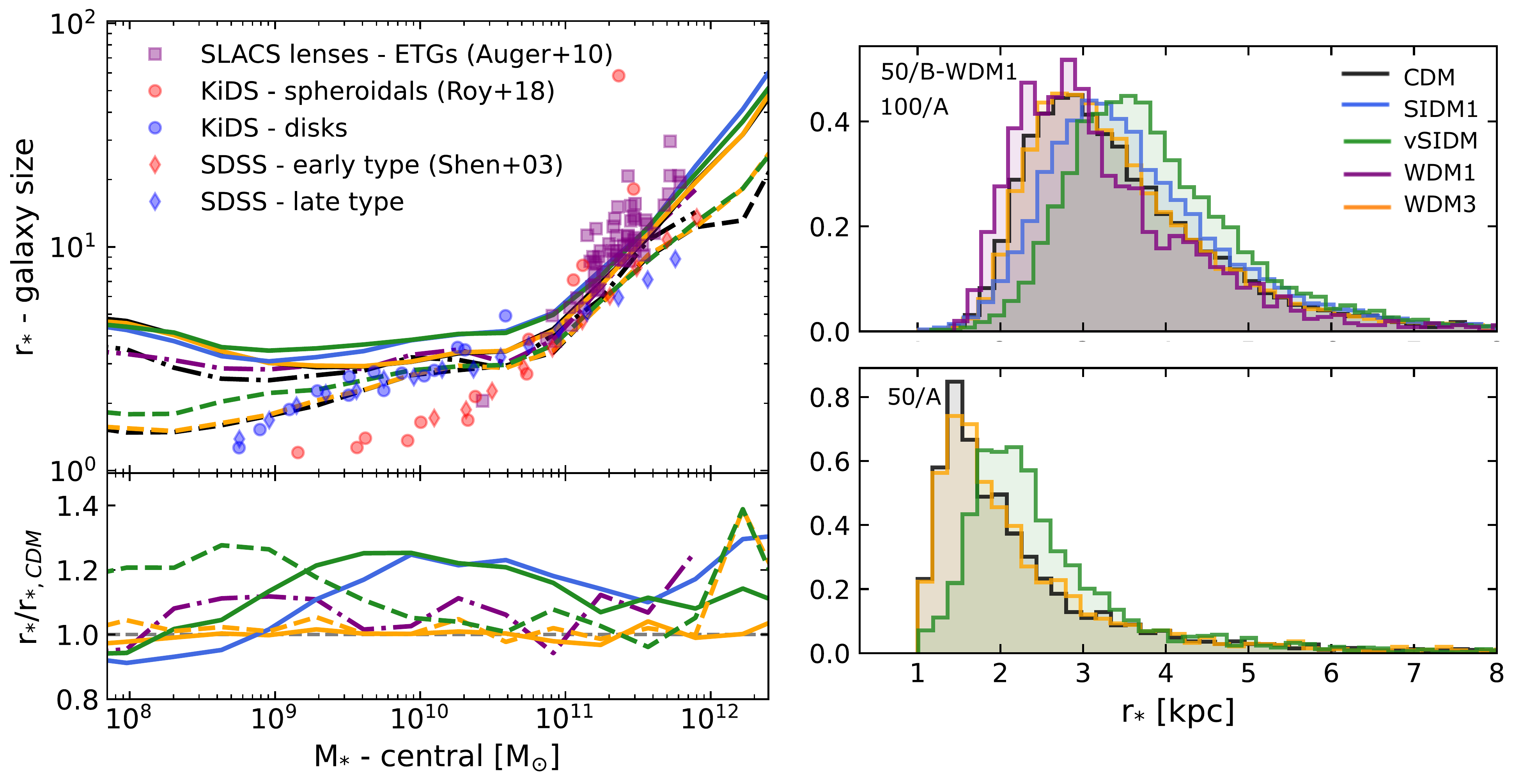}
   \caption{{\it Left:} Galaxy sizes expressed in terms of $r_{*}$ as a function of the galaxy stellar mass at $z=0$. The ratios of alternative models to CDM (bottom subpanel) show us that SIDM models produce larger galaxies in most mass bins (see also the bottom panel of Fig. \ref{fig:visual2} as an example). We show the running mean for the 50/A (dashed), 100/A (solid) and 50/B (dot-dashed) runs, the latter only to represent the WDM1 model. We compare simulated results to observational values from the SLACS strong gravitational lenses \citep[ETGs][]{auger10a}, galaxies (spheroidal and disk) from the KiDS survey \citep{roy18} and from the SDSS (early and late type) survey \citep{shen03}. We note, however, that the galaxy size is estimated as the projected half-light radius in observations while we plot the stellar half-mass radius, and the observed galaxies are at different redshifts. They are thus only meant to provide a qualitative comparison. {\it Right:} Full normalised distribution of sizes highlighting the lack of galaxies with small sizes in SIDM models.}
              \label{fig:gal_size}%
\end{figure*}

In the AIDA simulations, we employ the TNG galaxy formation model and leave it unchanged in all runs. Since the model was calibrated in a CDM universe, it is natural to ask whether or not alternative models may break it in some way and, in this case, how are the galaxy properties modified. In Fig.~\ref{fig:massf_st}, we have already shown that the stellar mass function at $z=0$ is not significantly affected. We now look at the other galaxy properties that have been considered in the calibration of the TNG model. In Fig.~\ref{fig:gal_prop}, we consider four of these properties related to the stellar and gas distribution in galaxies: the stellar mass fraction as a function of halo mass $M_{200c}$ at $z=0$ (top left), the gas fraction within $r_{500c}$ as a function of halo mass (top right), the SMBH mass as a function of stellar mass (bottom left), and the star-formation rate density (SFRD) as a function of redshift (bottom right). Despite the clear impact of ADM models on the halo mass function and density profiles, these global properties of galaxies are very similar in all considered scenarios. Compared to the results by \citet{pillepich18}, the difference between dark matter scenarios is much smaller than that due to variations of baryonic physics (see their Fig. B1). Galaxy properties are also affected by the resolution of the simulations, as one can appreciate in the panels showing multiple boxes. The difference between the 50/A and 100/A curves is again consistent with the analysis by \citet[][Fig. A2]{pillepich18} and \citet[][Fig. 2]{pakmor23}, while the effect of dark matter physics at fixed resolution induces much smaller changes. For reference, the resolution of the 50/A boxes is in between that of the TNG50-1 and TNG100-1 runs, while 100/A is comparable to TNG300-1. 

Figure~\ref{fig:gal_size} shows another observable property of galaxies that has been considered in the calibration: galaxy size (represented by the stellar half-mass radius) as a function of stellar mass. In the left panel, the running means from simulations are compared to observational measurements: spheroidal and disk galaxies from the KiDS survey \citep{roy18}, early and late types from the SDSS survey \citep{shen03}, and elliptical galaxies acting as strong gravitational lenses from the SLACS sample \citep[ETGs][]{auger10a}. Unlike Fig.~\ref{fig:gal_prop}, here we observe a systematic difference between CDM and SIDM: Self-interactions drive dark matter particles outwards from the halo centre, leading to a lower dark matter density and galaxy sizes larger by a factor of $\sim$20\% in the mass range between $5\times10^{9}$ and $10^{12}\,{\rm M}_{\odot}$. A similar trend is also present at higher redshift, maintaining the same distribution up to $z=0.5$ and then decreasing to $\sim$10\% at higher redshift. In the right panels, we show the normalised distribution of sizes separating the two levels of resolution to highlight the lack of systems with small $r_{*}$ in both cases. The velocity-dependent cross-section (green line) has the largest effect below $M_{*}=10^{10}\,{\rm M}_{\odot}$, while the constant cross-section impact is larger at higher masses. This difference can be identified in the example haloes shown in Fig.~\ref{fig:visual1}, where vSIDM produces the largest effective radius $r_{*}$, and Fig.~\ref{fig:visual2} where instead it is SIDM1. 

We conclude from the results presented in this section that there is a reassuring agreement between the AIDA simulations and the TNG original runs, meaning that most of the results found for CDM in TNG papers can also be expected to hold in all AIDA simulations. Despite the differences in the dark matter physics that emerges in the halo mass functions and profiles, all scenarios produce similar galaxy properties, indicating that the TNG galaxy formation model can produce a realistic population in all scenarios. This gives us a big advantage since we can use previous results to factor out baryonic effects and unveil those due to the physics of dark matter. Global properties such as those analysed in Fig.~\ref{fig:gal_prop} are not systematically modified, and it is thus not necessary to recalibrate the galaxy formation model at this stage to study the galaxy population in WDM and SIDM. However, properties that depend on the redistribution of matter within haloes due to dark matter physics are affected, such as the distribution of galaxy sizes shown in Fig.~\ref{fig:gal_size}. The visualisations in Fig.~\ref{fig:visual1}, \ref{fig:visual2}, and \ref{fig:visual3} indeed suggest that additional signatures can be found when looking at small scales. This will be the focus of upcoming AIDA papers.


\section{Matter power spectrum} \label{sec:power}
\begin{figure}

   \includegraphics[width=\linewidth]{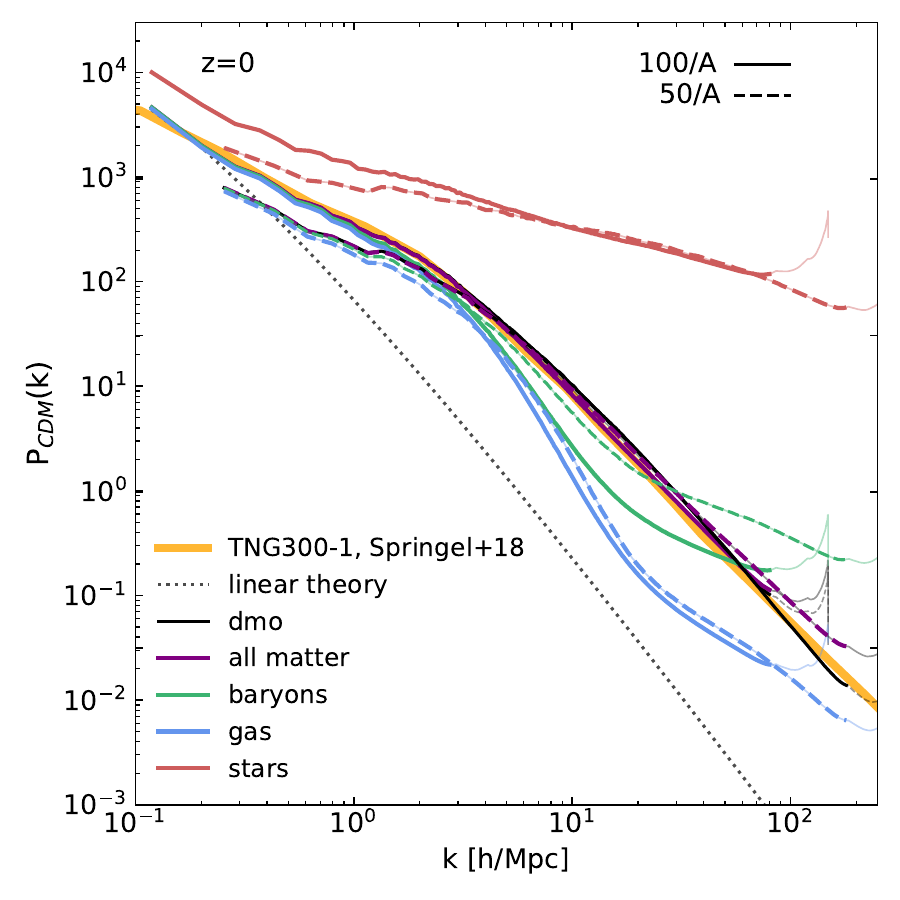}
   \caption{Matter power spectrum in the CDM runs at $z=0$ for the DMO (black) and the FP runs separated into different components (coloured curves). The dotted lines show the prediction from linear theory, while the yellow line shows the result from \citet{springel18} for the larger TNG300-1 box. We used the two high-resolution boxes, 100/A (solid lines) and 50/A (dashed lines). All power spectra are computed with a grid of size $2048^{3}$, leading to Nyquist frequencies for the two cases of $k=85.8\,h\,{\rm Mpc}^{-1}$ and $k=183\,h\,{\rm Mpc}^{-1}$, respectively. Here and in the following figures, we used these values to limit the curves and plot the remaining points as a fainter line.
   }
              \label{fig:Pk1}%
\end{figure}

In this section we investigate the combined effect of baryons and ADM on the matter power spectrum and its evolution with redshift. We already know that the initial power spectrum is suppressed in WDM models (see Fig. \ref{fig:models}) and that we expect some of this suppression to persist until $z=0$, given that the number density of low-mass haloes is lower than in CDM (see Fig.~\ref{fig:massf}). Instead,  the initial power is unchanged in SIDM, and so any difference compared to CDM is created by self-interactions during the evolution of structures. At the same time, baryonic physics alters the power on scales smaller than $\sim$1 Mpc \citep{hellwing16,springel18}. We thus expect that baryonic and dark matter effects will combine at these scales, where they both appear. We note that in this Sect. we express distances in units of Mpc $h^{-1}$ and kpc $h^{-1}$ to match the usual convention of the field.

We calculated the matter power spectrum in each box by sampling the mass distribution with a $2048^{3}$ grid. In the FP simulations, we computed the total matter power spectrum as well as the spectrum of each mass component separately. Figure~\ref{fig:Pk1} shows the relative power of dark matter and baryons, separated in gas and stars, in the CDM boxes at $z=0$. In the same figure, we also show the DMO spectrum, the prediction from linear theory and the results from \citet{springel18} -- with which we find very good consistency.  We compare the two high-resolution boxes to each other (solid versus dashed lines). As expected, the smaller 50/A boxes cannot sample the power spectrum correctly on large scales, while the 100/A runs follow well the results from \citet{springel18}. The situation is reversed at small scales, and the measurements agree well in the intermediate range covered by both boxes.

\begin{figure*}
    \centering
   \includegraphics[width=\linewidth]{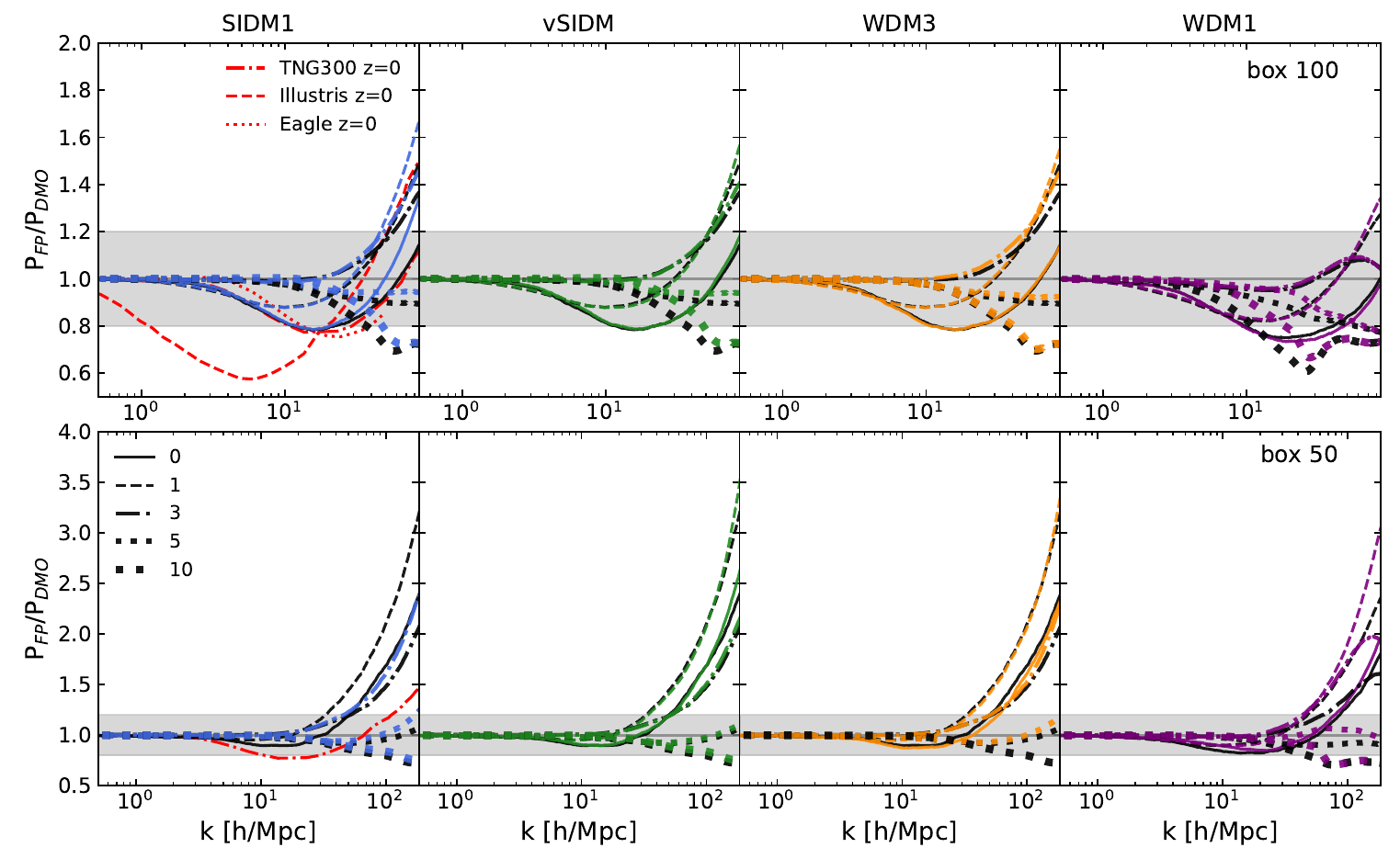}
   \caption{Baryonic effects on the total matter power-spectrum measured at five different redshifts from $z=0$ to $z=10$. Here we plot the $P(k)$ ratio of the FP run of each model to its corresponding DMO one. Each column considers one alternative model (coloured lines of different styles) and contrasts it with the CDM result at the same redshift (black lines). In the top (bottom) panels, we look at the 100/A (50/A) boxes. We see how the smaller box allows us to reach higher $k$ values and larger enhancements of the power spectrum due to the baryons located at the very centre of haloes. The power spectrum of the WDM1 model (rightmost panels) is measured in the 100/B and 50/B boxes and compared to the corresponding CDM case -- the reason why the black curves differ in the right panel. The grey bands mark a difference within 20\% from the DMO value. In the leftmost panels, we also compare to previous results at $z=0$ from the Illustris, TNG300 \citep{springel18} and Eagle \citep{hellwing16} simulations.
   }
              \label{fig:Pkbar}%
\end{figure*}
\begin{figure*}
   \includegraphics[width=0.95\linewidth]{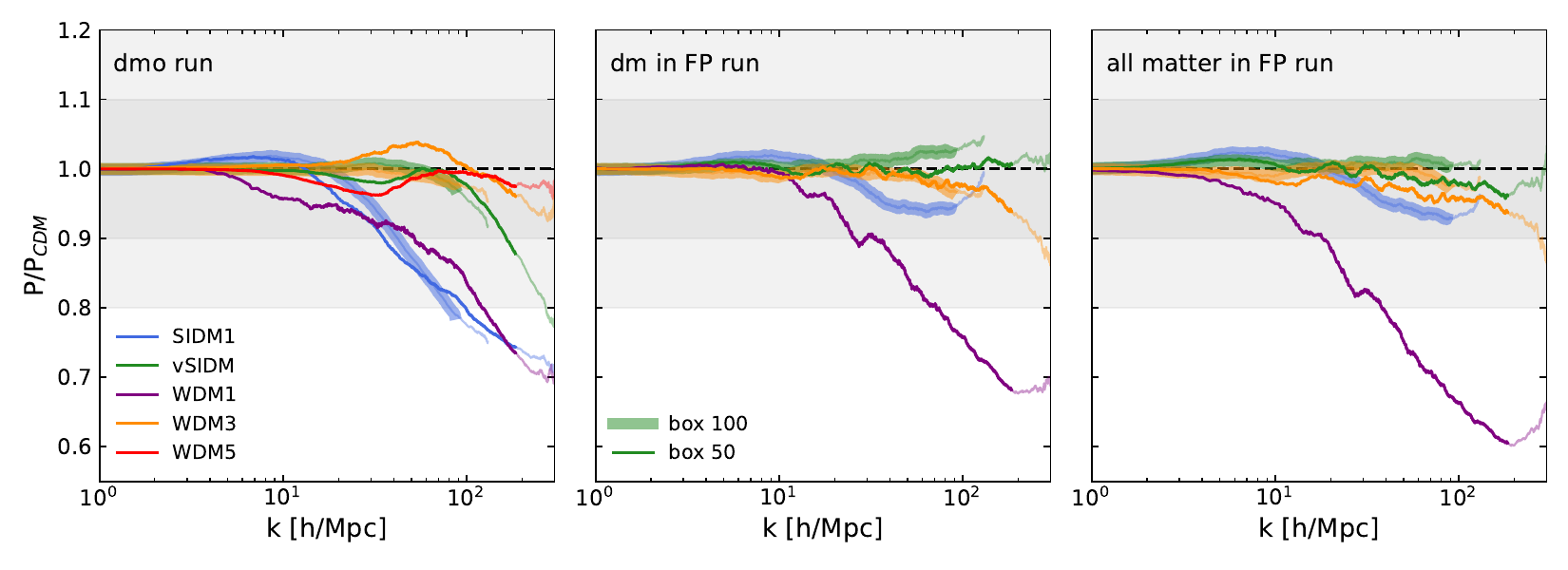}
   \caption{Matter power spectrum in alternative models compared to the CDM scenario in the dark (left) and FP (middle and right) runs at $z=0$. For each dark matter model, we plot the ratio to the CDM value in the range $1\, h\,{\rm Mpc}^{-1}\leq k\leq 300\, h\,{\rm Mpc}^{-1}$, corresponding to scales $1\,h^{-1}{\rm Mpc}\geq l\geq0.003\,h^{-1}{\rm Mpc}$. At scales larger than $1\,h^{-1}{\rm Mpc}$, the power spectra are identical, indicating that the large-scale structure distribution is unaltered by ADM effects. Each dark matter model is represented by a different colour, while the thin (thick) lines show the results for the 50 (100) Mpc boxes. }
              \label{fig:Pk_z0}%
\end{figure*}
\begin{figure*}
    \centering
   \includegraphics[width=0.91\linewidth]{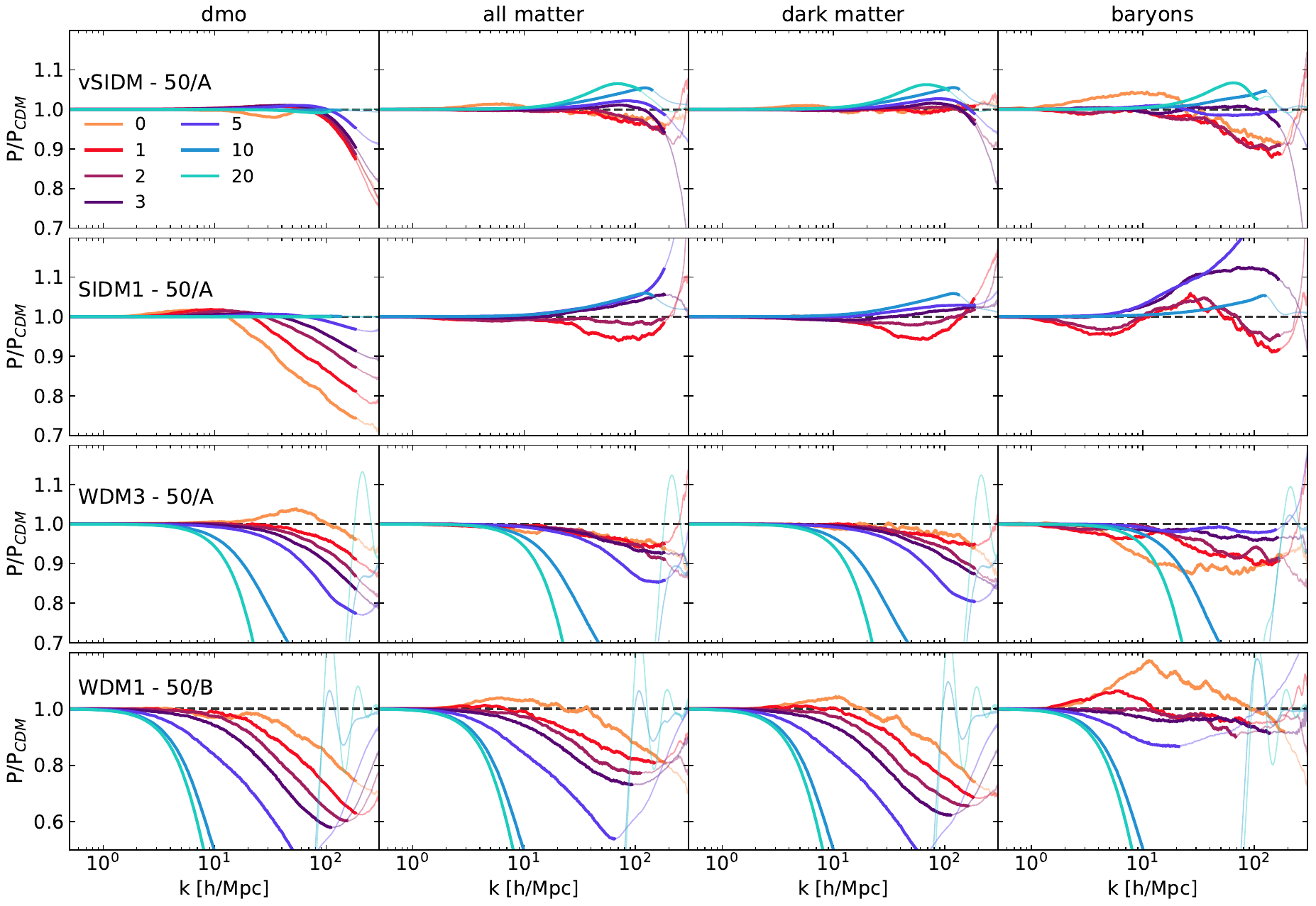}
   \includegraphics[width=0.92\linewidth]{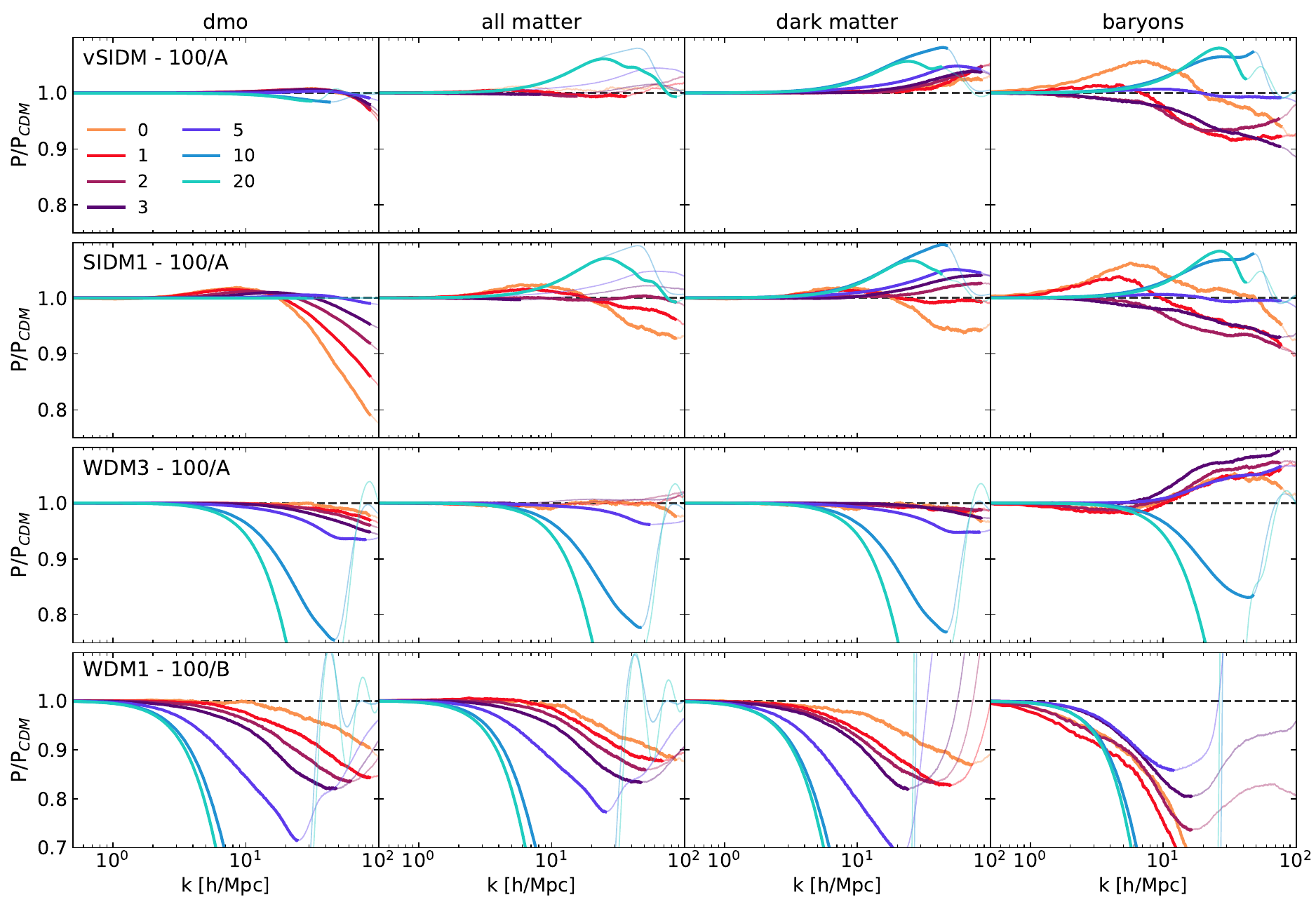}
   \caption{Redshift evolution of the matter power spectra in alternative models compared to the CDM scenario in the two boxes. Each panel shows the ratio between the two for a particular case. From left to right, we show the DMO power and the total power in the FP, followed by the separate power spectra of dark matter and baryons in the FP boxes. Each colour stands for a redshift from $z=20$ (light blue) to $z=0$ (orange).}
              \label{fig:allcomp1}%
\end{figure*}

\subsection{Baryonic effects}

We first look at the effect of baryons in each dark matter model. To this end, we compared the total matter power spectrum in the FP runs to the corresponding DMO one. In Fig.~\ref{fig:Pkbar}, we visualise this as $P_{\rm FP}/P_{\rm dmo}$ ratios measured at five redshifts -- $z=0$, 1, 3, 5, and 10 -- in the 100 (top) and 50 (bottom) Mpc boxes. The CDM results are shown in black and are identical in all panels, while the other colours stand for the four alternative models for which we have hydrodynamical runs. Baryons start suppressing the power spectrum at $k\geq1\, h\,{\rm Mpc}^{-1}$ and have the strongest effect at $z=0$ and $k\sim15\, h\,{\rm Mpc}^{-1}$, corresponding to $r\sim100$ $h^{-1}$kpc. At smaller scales, close to the centre of galaxies and clusters, the power is instead rapidly enhanced in the FP simulations. Interestingly, baryonic effects have an almost identical effect on the power spectrum of all dark matter models.
This indicates that the drastic changes in the baryon distribution relative to the dark matter brought about by galaxy formation physics are a dominant factor in determining the change of the total matter power spectrum, despite differences in the underlying dark matter physics. 

To aid the comparison to previous works, we also show measurements from the Illustris, TNG300 \citep{springel18} and Eagle \citep{hellwing16} simulations. As already discussed in \citet{springel18}, the effect is noticeably stronger in Illustris than in all other runs: The suppression extends to considerably larger scales and is stronger in amplitude due to the important differences in the AGN feedback model. On the other hand, the TNG300 and Eagle measurements are close to the CDM measurement (black solid line) presented here.

Some small differences are visible at  $k\geq50\, h\,{\rm Mpc}^{-1}$, corresponding to scales $r\leq 20\,h^{-1}\,{\rm kpc}$: in self-interacting scenarios, the power spectrum is enhanced compared to CDM at $z<3$, with SIDM1 showing the strongest effect, followed by vSIDM. This is consistent with the fact that SIDM1 (followed by vSIDM) haloes have the strongest response to the presence of baryons at the centre of haloes and to adiabatic contraction, which led to much denser central regions compared to the dark runs -- see the middle panels of Fig.~\ref{fig:prof1}. This effect is only visible at low redshift ($z\leq1$), when the effect of self-interactions strongly kicks in.

\subsection{The impact of dark matter physics}

Next, we looked into the effect of ADM at $z=0$. In Fig.~\ref{fig:Pk_z0}, we plot the ratio between the ADM power spectra and the CDM one. The power is unchanged for $k<1\, h\,{\rm Mpc}^{-1}$, corresponding to length scales larger than $1\,h^{-1}\,{\rm Mpc}$. The large-scale distribution of structures which unfolds in the cosmic web of filaments, walls, and knots is similar in all models. In the DMO runs (left panel), the power is suppressed in all alternative models, although by different amounts. The two boxes show a good convergence: The 50 Mpc and 100 Mpc boxes are represented, respectively, by the thin and thick (fainter) lines. The SIDM1 model shows a very strong difference from CDM, reaching a level that is comparable to the most extreme WDM1 model at high-$k$. However, WDM1 starts departing from CDM at larger scales ($k\simeq4$ instead of $k\simeq20$), thus showing the different origin of the effect. Warm models show a global effect due to the lack of low-mass systems, while self-interactions appear only at the very centre of haloes. 

The two colder WDM models are more similar to CDM, indicating that the suppression introduced in the initial conditions has been largely reduced at $z=0$. The other panels of Fig.~\ref{fig:Pk_z0} show measurements from the FP runs, using only the dark matter component (middle) and all matter (right). The middle panel thus illustrates how much the presence of baryons impacts the dark matter distribution. Different models react differently to the presence of baryons on small scales: In WDM1 and WDM3, the suppression is still clear, while the SIDM1 and vSIDM power spectra are now much more similar to than in the dark runs. The total matter power spectra (right panel) show a similar behaviour. This can be understood in terms of the different physical origins of the suppression: Given that baryons cool and form galaxies at the centres of the available dark matter (sub)haloes, their presence does not increase the number of low-mass haloes and subhaloes in WDM models (see the halo mass functions in Fig. \ref{fig:massf}). On the other hand, when such baryonic structures are formed at the centres of haloes, they reduce the fraction of dark matter in the inner regions and, with it, the fraction of matter that is affected by self-interactions; their presence prevents the formation of large cores in the total density profiles which remain closer to the CDM case (see Fig.~\ref{fig:prof2}).

The AIDA runs allow us to follow the evolution of $P(k)$ throughout the entire history of the Universe and determine the redshifts at which WDM and SIDM effects are most important. Moreover, we can also follow each component (dark matter, stars and gas) separately to understand how the interplay between dark matter physics and galaxy formation impacts each of them. Figure~\ref{fig:allcomp1} focuses on the redshift evolution of the power spectrum of different components in each dark matter scenario, computed at six representative redshifts between $z=20$ and $z=0$. To highlight the differences to CDM at small scales, we plot again the ratios P/P$_{\rm CDM}$ for the dark runs (first column) and different components of the FP runs (other columns). In the DMO runs, WDM shows, by construction, the strongest reduction of power on small scales at high-$z$ due to the lack of low-mass haloes and subhaloes, and the formation of cores in haloes with masses $M_{\rm vir}\leq0.1\times M_{\rm hm}$. The relative difference to CDM is then mitigated towards $z=0$ when structure formation becomes increasingly more non-linear. Instead, SIDM models evolve in the opposite direction: They are equivalent to CDM at high-$z$, while differences emerge later when the self-interactions create large cores at the centre of haloes (see Fig.~\ref{fig:prof1}, which are the main source of the lack of power at $k>10\,h\, {\rm Mpc}^{-1}$ ($l<100 \, h^{-1} {\rm kpc}$). By $z=0$ the suppression in SIDM1 is comparable to or stronger than in WDM models: in the SIDM1 model, we have 10\% less power at $k=20\,h\, {\rm Mpc}^{-1}$ and 30\% at $k>200\,h\, {\rm Mpc}^{-1}$. The velocity-dependent model is more complex, as the self-interaction cross-section is inversely proportional to the relative velocity of particles, which is a proxy for halo mass. The vSIDM cores are smaller than in SIDM1 at the high-mass end (see Fig.~\ref{fig:prof1}) and they disappear for $M_{\rm vir}\leq10^{12}\,{\rm M}_{\odot}$. This is reflected in a weaker suppression that is relevant at $k\geq60\, h\,{\rm Mpc}$ ($l\leq20\,h^{-1}{\rm kpc}$). 

The second column in Fig.~\ref{fig:allcomp1} shows the ratio of total matter power spectra measured in the full physics runs, calculated with all particle types (dark matter, gas, stars and SMBHs). In all scenarios, the $z=0$ suppression is less pronounced than in the DMO runs due to the presence of baryonic matter on small scales, which is unaffected by dark matter physics. In both WDM models, this is the case at all redshifts, while SIDM produces a more complex evolution with fluctuations within 10\% around P/P$_{\rm CDM}=1$. The next two columns divide the FP runs into dark and baryonic matter to show the relative contributions of the two. For example, from Fig.~\ref{fig:Pk1}, we know that stars dominate the power at all scales compared to gas. Unexpected variations of baryonic power can thus be related to differences in the star formation history of the simulation. In the bottom-right panel for Fig. \ref{fig:gal_prop}, the SFRD is higher in WDM3 than in CDM for $z>2.5$, which is related to the higher power in the baryonic component in this model. 

\section{Summary and conclusions} \label{sec:conc}

This work introduces the AIDA-TNG project (AIDA for short hereafter) and presents the first results from two cosmological volumes of 110.7 and 51.7 Mpc on a side, corresponding to 75 and $35\,h^{-1}{\rm Mpc}$. Each volume is evolved to $z=0$ in multiple dark matter scenarios, including a complete treatment of baryonic physics, described by the IllustrisTNG galaxy formation model \citep{weinberger17,pillepich18}. Apart from CDM, we consider three WDM models with particle masses $m_{\rm WDM}= (1,3,5)\,{\rm keV}$, and two self-interacting dark matter scenarios where the cross-section is either constant ($\sigma=1\,{\rm cm^{2}g^{-1}}$) or follows the velocity-dependent model by \citet{correa21}. Our runs reach a maximum mass resolution of $m_{\rm DM}=3.6\times10^{6}\,{\rm M}_{\odot}$ and $m_{\rm bar}=6.8\times10^{5}\,{\rm M}_{\odot}$ in our 50/A flagship run. The family of runs of AIDA-TNG are listed and summarised in Table~\ref{table:1}. These will be complemented by higher-resolution 20 Mpc box, which will be presented in an upcoming work.

We present the first analysis of the AIDA runs in terms of halo and stellar mass functions, density profiles, galaxy properties and matter power spectra. We compare the outcome of the full physics runs (FP) with those including only dark matter (DMO) and across dark matter models. These comparisons reveal the scales that are most affected by the physics of dark matter and the somewhat opposite trends in WDM and SIDM.  

In Sect.~\ref{sec:massf}, we measured the halo mass function at redshifts between $z=0$ and $z=5$ and quantified the effects of the dark matter models and baryons. We recovered the suppression at small scales predicted by WDM and found good consistency with previous results. Self-interactions do not modify the mass function in the DMO runs, while in the FP ones they produce a $\sim5-10$\% increase in the number density of haloes below $10^{11}\,{\rm M}_{\odot}$ when compared to CDM. In Sect.~\ref{sec:prof}, we measured the dark matter and total density profiles of haloes over six orders of magnitudes in mass -- from $10^{9}\,{\rm M}_{\odot}$ to $10^{14}\,{\rm M}_{\odot}$ -- and extract the halo concentration-mass relation. In the DMO runs, self-interactions create density cores of sizes increasing with mass, leading to a very different dark matter distribution in haloes. Instead, WDM creates cores in low-mass haloes, starting at a mass about 100 times the model half-mode mass. These trends persist in the FP runs, although the presence of baryons reduces the sizes and depth of dark matter cores. Moreover, SIDM models produce steeper inner slopes than CDM at intermediate masses $M_{\rm vir}\leq10^{12}\,{\rm M}_{\odot}$. These signatures of warm and self-interacting dark matter are visible not only in the dark matter density but also in the stellar and total matter distribution, highlighting the interplay between dark and luminous matter.

The design of the AIDA project also allows us to perform quantitative tests of the galaxy formation model in all dark matter scenarios and evaluate the properties of the galaxy population. In Sect.~\ref{sec:galaxies}, we move to the properties of galaxies and perform quantitative tests of the galaxy formation model in all considered scenarios. We look at quantities that have been considered in the design of the TNG model. The $z=0$ stellar and gas mass fractions within haloes, the $z=0$ super-massive black hole mass, and the cosmic star formation rate density are all well consistent across dark matter scenarios. The distribution of galaxy sizes instead shows a signature of self-interactions, which create larger sizes compared to CDM. This indicates that alternative models considered here affect properties that depend on the distribution of matter within haloes and galaxies, but not scaling relations associated with the total halo or stellar mass. Similarly, in Sect.~\ref{sec:massf}, we found that the stellar mass function is not significantly different across models, and thus, the prediction of the model for galaxy counts is preserved. 
Moreover, we compared the FP and DMO of each scenario and found that the TNG galaxy formation model produces a similar suppression of the halo mass function (Sect.~\ref{sec:massf}) and the matter power-spectrum (Sect.~\ref{sec:power}) in all models with some deviations only in the WDM1 runs at redshift $z=2$ and above. 
These results suggest that the TNG model can reproduce the overall properties of structures and galaxies in all considered scenarios and that baryonic effects can, to first order, be factored out when comparing dark matter models. Moreover, compared to the analysis by \citet{pillepich18}, who considered variations in the baryonic model, we see that the latter introduces much larger differences in galaxies compared to ADM. Nevertheless, some properties of galaxies are affected by the physics of dark matter: The total and stellar density profiles are $(i)$ cored at high masses in SIDM, $(ii)$ cuspy at the scale of massive galaxies and below in SIDM, and $(iii)$ are cored at low masses in WDM. Moreover, the stellar distribution in SIDM galaxies is more extended than in CDM. 

It is well known that baryons also modify the matter power spectrum on scales smaller than 1~Mpc, creating a suppression of $\sim20$\%. In Sect.~\ref{sec:power} we find that this is once again very similar across all dark matter models despite the larger differences between alternative models and CDM.  We also quantify the variations in the small-scale power due to ADM physics, considering the dark matter, total and baryonic power spectra. Once again, WDM and SIDM models show opposite trends: The initial WDM suppression is reduced towards $z=0$, while self-interactions create important variations towards low redshift.

The AIDA-TNG project offers a unique opportunity to study structure and galaxy formation in ADM scenarios. In this work, we showed that it can be used to generate predictions at multiple scales and a variety of cosmological environments while simultaneously capturing subgalactic physics. New and upcoming observational facilities, such as Euclid, JWST, Vera Rubin, ALMA, VLBI and ELT will soon provide large statistical samples of galaxies and exciting progress in resolution, giving us a sharper view of the properties of dark matter. AIDA is extremely timely: Our simulations will be instrumental in guiding the interpretation of existing and upcoming observations, giving us a fantastic window of opportunity to make an impact in the field of dark matter and galaxy formation.

\section*{Data availability}

The original IllustrisTNG simulations are publicly available and accessible at \url{www.tng-project.org/data} \citep{nelson19}, where AIDA-TNG will also be made public in the future. Data directly related to this publication is available on request from the corresponding author. Additional information about AIDA can be found online at \url{https://gdespali.github.io/AIDA/}.

\begin{acknowledgements}
GD thanks Haibo Yu, Ruediger Pakmor, Daneng Yang and Carlo Giocoli for useful discussions on the models and the code. GD would like to thank her late PhD supervisor Bepi Tormen for teaching her how to run simulations and always find the best names for them. We acknowledge the EuroHPC Joint Undertaking for awarding this project access to the EuroHPC supercomputer LUMI, hosted by CSC (Finland) and the LUMI consortium through a EuroHPC Extreme Scale Access call. We acknowledge ISCRA and ICSC for awarding this project access to the LEONARDO supercomputer, owned by the EuroHPC Joint Undertaking, hosted by CINECA (Italy). GD acknowledges the funding by the European Union - NextGenerationEU, in the framework of the HPC project – “National Centre for HPC, Big Data and Quantum Computing” (PNRR - M4C2 - I1.4 - CN00000013 – CUP J33C22001170001). LM acknowledges the financial contribution from the PRIN-MUR 2022 20227RNLY3 grant “The concordance cosmological model: stress-tests with galaxy clusters” supported by Next Generation EU and from the grant ASI n. 2024-10-HH.0 “Attività scientifiche per la missione Euclid – fase E”. DN acknowledges funding from the Deutsche Forschungsgemeinschaft (DFG) through an Emmy Noether Research Group (grant number NE 2441/1-1). We thank the referee for the appreciation of our work.

\\This research made use of the public Python packages matplotlib \citep{matplotlib}, NumPy \citep{numpy}, \textsc{COLOSSUS} \citep{diemer18} and Py-SPHViewer \citep{py-sphviewer}.

\end{acknowledgements}

\bibliographystyle{aa}
\bibliography{aa53836-25corr.bbl}

\end{document}